\documentclass[review,3p]{elsarticle}

\usepackage{graphicx}

\usepackage[breaklinks]{hyperref}

\graphicspath{%
  {./figures/}}

\usepackage{amssymb}

\usepackage[pagewise,modulo]{lineno}

\usepackage{alphalph}
\renewcommand{\theaffn}{\alphalph{\value{affn}}}
\newcommand{\vmet}{\ensuremath{\vec{\slash\kern-.7emE}_{T}}}
\newcommand{\ppbar}{\ensuremath{p\overline{p}}}
\newcommand{\zee}{\ensuremath{Z \rightarrow ee}}
\newcommand{\wen}{\ensuremath{W \rightarrow e \nu}}

\journal{Nuclear Instruments and Methods A}

\title{A novel method for modeling the recoil in $W$ boson events at hadron colliders}

\begin{document}

\begin{frontmatter}
%
\author[aff37]{V.M.~Abazov}
\author[aff75]{B.~Abbott}
\author[aff65]{M.~Abolins}
\author[aff30]{B.S.~Acharya}
\author[aff51]{M.~Adams}
\author[aff49]{T.~Adams}
\author[aff6]{E.~Aguilo}
\author[aff59]{M.~Ahsan}
\author[aff37]{G.D.~Alexeev}
\author[aff41]{G.~Alkhazov}
% ...

\address{\vspace{0.1 in}(The D\O\ Collaboration)\vspace{0.1 in}}
% ...
\address[aff6]{University of Alberta, Edmonton, Alberta, Canada;
                Simon Fraser University, Burnaby, British Columbia, Canada;
                York University, Toronto, Ontario, Canada and
                McGill University, Montreal, Quebec, Canada}
% ...
\address[aff30]{Tata Institute of Fundamental Research, Mumbai, India}
% ...
\address[aff37]{Joint Institute for Nuclear Research, Dubna, Russia}
% ...
\address[aff41]{Petersburg Nuclear Physics Institute,
                St. Petersburg, Russia}
% ...
\address[aff49]{Florida State University, Tallahassee, Florida 32306, USA}
% ...
\address[aff51]{University of Illinois at Chicago,
                Chicago, Illinois 60607, USA}
% ...
\address[aff59]{Kansas State University, Manhattan, Kansas 66506, USA}
% ...
\address[aff65]{Michigan State University,
                East Lansing, Michigan 48824, USA}
% ...
\address[aff75]{University of Oklahoma, Norman, Oklahoma 73019, USA}
% ...

\fntext[fna] {
Visitor from Augustana College, Sioux Falls, SD, USA.}
\fntext[fnb] {
Visitor from Rutgers University, Piscataway, NJ, USA.}
\fntext[fnc] {
Visitor from The University of Liverpool, Liverpool, UK.}
\fntext[fnd] {
Visitor from Centro de Investigacion en Computacion - IPN,
  Mexico City, Mexico.}
\fntext[fne] {
Visitor from ECFM, Universidad Autonoma de Sinaloa, Culiac\'an, Mexico.}
\fntext[fnf] {
Visitor from Universit{\"a}t Bern, Bern, Switzerland.}
\fntext[fng] {
Visitor from Universit{\"a}t Z{\"u}rich, Z{\"u}rich, Switzerland.}

  \begin{abstract}
We present a new method for modeling the hadronic recoil
in $W\rightarrow \ell \nu$ events produced at hadron colliders.
The recoil is chosen from a library of recoils in $Z\rightarrow \ell \ell$
data events and overlaid on a simulated $W \rightarrow \ell \nu$ event. 
Implementation of this method requires that the data recoil library 
describe the properties of the measured recoil as a function of the true, 
rather than the measured, transverse momentum of the boson.
We address this issue using a multidimensional Bayesian
unfolding technique. We estimate the statistical and systematic
uncertainties from this method for the $W$ boson mass and width measurements
assuming 1 fb$^{-1}$ of data from the Fermilab Tevatron. The uncertainties are
found to be small and comparable to those of a more traditional
parameterized recoil model. For the high precision measurements
that will be possible with data from Run II of the Fermilab Tevatron and from the CERN LHC,
the method presented in this paper may be advantageous,
since it does not require an understanding of the measured recoil from first principles.
  \end{abstract}

\end{frontmatter}


\section{Introduction}
\label{sec-intro}
The $W$ and $Z$ bosons are massive gauge bosons that, along with 
the massless photon, mediate electroweak interactions. 
The predictions from the standard model (SM) of weak, electromagnetic,
and strong interactions~\cite{bib:sm} for
their masses and widths include radiative corrections from the top quark and 
the Higgs boson. When precision measurements of the $W$ boson mass ($M_W$) are 
combined with measurements of the top quark mass and other electroweak 
observables, limits on the Higgs boson mass can be extracted. 
The $W$ boson width ($\Gamma_W$) can be directly measured 
from the fraction of $W$ bosons produced at high mass. It can also be inferred 
indirectly within the context of the SM from the leptonic branching fraction of 
the $W$ boson. The branching fraction, in turn, can be inferred from the ratio of the
$W$ and $Z$ boson cross-sections with additional theoretical inputs~\cite{bib:width_indirect}.
The direct measurement of $\Gamma_W$ is sensitive to vertex 
corrections from physics beyond the SM.  
The current world average for $M_W$ is $80.398 \pm 0.025$ GeV 
\cite{cdfw} and the current world average for $\Gamma_W$ is 
$2.106 \pm 0.050$ GeV from direct measurements~\cite{bib:currentWidth}. 
The large number of $W$ bosons currently available
in data samples collected at the Fermilab Tevatron collider and 
that will soon be available from the CERN LHC collider allow measurements 
of $M_W$ and $\Gamma_W$ with unprecedented precision
provided the response of the detector can be modeled with sufficient accuracy.

\begin{figure}[htbp]
  \begin{center}
    \includegraphics[scale=0.6]{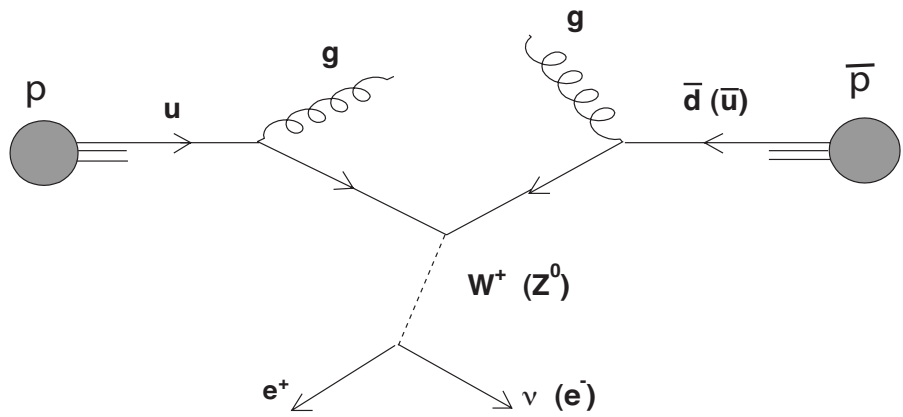}
  \end{center}
  \caption{An example of a diagram for the production and leptonic decay of a 
           $W$/$Z$ boson with radiated gluons in a hadronic collision.}
  \label{fig:feyn}
\end{figure}

In $p\bar{p}$ and $pp$ collisions, $W$ and $Z$ bosons are produced predominantly
through quark-antiquark annihilation. Higher order processes can include 
radiated gluons or quarks that recoil against the boson
and introduce non-zero boson transverse momentum~\cite{bib:resbos}.
Figure~\ref{fig:feyn} shows an example diagram for the production 
of a $W$ or $Z$ boson with two radiated gluons in a $p\bar{p}$ collision.

We identify $W$ and $Z$ bosons through their leptonic decays 
($W \rightarrow \ell\nu$ and $Z\rightarrow \ell \ell$ with 
$\ell=e, \mu$) since these signatures have low backgrounds. 
The charged leptons can be detected by the calorimeter or the muon system, 
while the neutrino escapes undetected. We do not reconstruct particles 
whose momentum vectors are nearly parallel to the beam direction, and therefore 
we only use kinematic variables in the transverse plane that is perpendicular to 
the beam direction. The neutrino transverse momentum vector ($\vec{p}_T^{~\nu}$) is 
inferred from the missing transverse energy ($\vmet$), which can be calculated 
using the transverse momenta of the charged lepton ($\vec{p}_T^{~\ell}$) and the 
recoil system ($\vec{u}_T$):
\begin{equation}
  \vmet = -[\vec{p}_T^{~\ell} + \vec{u}_T].
\end{equation} 
We measure $\vec{u}_T$ by summing the observed transverse energy 
vectorially over all calorimeter cells that are not associated with
the reconstructed charged lepton. 

\begin{figure*}[htbp]
  \begin{center}
    \includegraphics[scale=0.5]{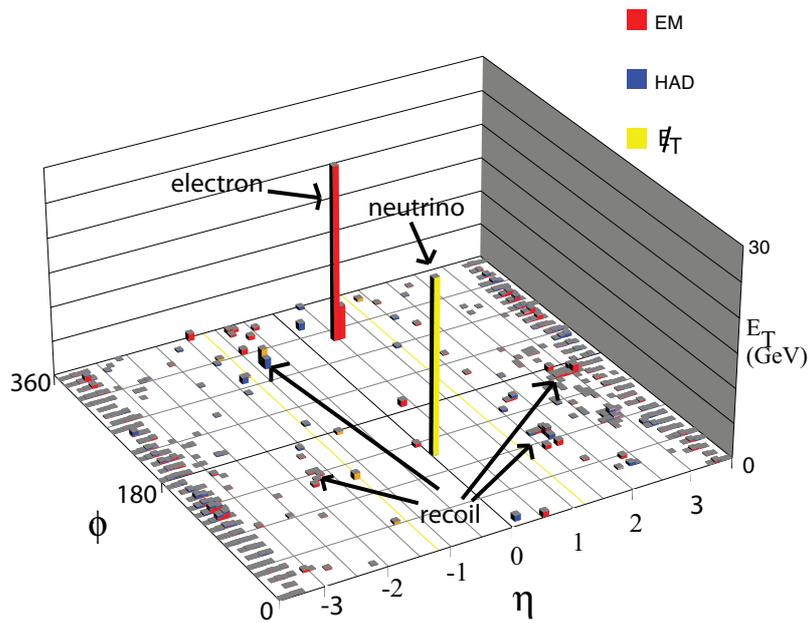}
  \end{center}
  \caption{A typical $W\rightarrow e \nu$ candidate as recorded by the D0 detector.
The two horizontal axes correspond to azimuthal angle and pseudorapidity,
and the vertical axis is the transverse energy deposited at that location in
the calorimeter. The energy associated with the electron and the ~\vmet~that 
corresponds to the neutrino are indicated. All other energies contribute to 
the measured recoil. The longitudinal component of the neutrino momentum 
is not determined, so it is displayed arbitrarily at $\eta=0$.}
  \label{fig:eventdisplay}
\end{figure*}

The recoil system is difficult to model from 
first principles; unlike the decay lepton, it is a complicated
quantity involving many particles, as well as effects related to 
accelerator and detector operation.
The recoil system is a mixture of the ``hard'' recoil that balances the boson 
transverse momentum and ``soft'' contributions, such as particles 
produced by the spectator quarks (the ``underlying event''), other
$p\bar{p}$ ($pp$) collisions in the same bunch crossing, electronics noise, 
and residual energy in the detector from previous bunch crossings (``pileup'').
Figure ~\ref{fig:eventdisplay} shows transverse energies recorded
in the calorimeter of the D0 detector versus azimuthal angle and 
pseudorapidity~\cite{bib:geo} for a typical $W \rightarrow e \nu$ candidate.  
The diffuse energy deposits spread over much of the detector are due to the recoil system.

The various components of this measured recoil system have different dependences 
on instantaneous luminosity. For example, pileup and additional inelastic collisions 
scale with instantaneous luminosity, while the contribution from the underlying event is 
luminosity independent. Moreover, detector effects such as 
suppression of calorimeter cells with low energy to minimize the event size
(zero-suppression cuts) can introduce correlations between the calorimeter 
response to the hard component and various soft components in the event, so that the 
detector responses to these components cannot be modeled independently.

Two approaches have been used previously to model the $W$ boson event, 
including the recoil system. One method takes the underlying physics from 
a standard Monte Carlo (MC) event generator and smears it parametrically to 
reproduce detector effects~\cite{{cdfw}, {review},{d0w},{cdfw2}}. The parameters 
are tuned to an independent but kinematically similar data set, namely 
$Z \rightarrow \ell \ell$ events. The second approach, or ``Ratio Method'', 
constructs $M_T$ template distributions by directly taking $Z\rightarrow \ell \ell$ 
events from collider data, treating one of the leptons as a neutrino~\cite{bib:Shpakov}. 
The ratio of the $Z$ boson mass to the corresponding $W$ boson mass, taken together with the 
precisely measured $Z$ boson mass~\cite{bib:zmass} from the CERN LEP collider determines the $W$ boson mass. 
In this method, small differences in the $Z$ and $W$ boson line shapes and transverse 
momentum and rapidity distributions of the decay leptons must be taken into account.

This paper presents a novel approach for modeling the recoil system in 
$W \rightarrow \ell \nu$ events at hadron colliders that uses 
recoils extracted directly from $Z \rightarrow \ell \ell$ collider 
data. The $Z \rightarrow \ell \ell$ data provide a mapping between the $Z$ 
boson transverse momentum ($\vec{p}_T^{~Z}$) and the 
transverse momentum of the recoil system ($\vec{u}_T$).
Versions of the recoil library approach have been proposed in the 
past~\cite{bib:Saltzberg} that used the map between the reconstructed 
$\vec{p}_T^{~Z}$ and the measured $\vec{u}_T$ directly. In this paper we use a two-dimensional 
Bayesian unfolding method to derive a relation between the true 
$\vec{p}_T^{~Z}$ and the measured $\vec{u}_T$, which allows the simulation of the 
recoil system for the same generator level value of the $W$
boson transverse momentum ($\vec{p}_T^W$). 

The recoil library method presented in this paper has many advantages. 
Since the recoils are taken directly from $Z \rightarrow \ell \ell$ data, 
they reflect the event-by-event response and resolution of the detector. 
The additional soft recoil is built in, as is the complicated 
zero-suppression-induced correlations between it and the hard component 
of the recoil. Proper scaling of the recoil system with instantaneous luminosity is automatic 
since the $W$ and $Z$ samples have similar instantaneous luminosity profile. The most 
significant advantage of this method lies in its simplicity since
it does not require a first-principles understanding of the recoil system and
has no adjustable parameters. The dominant
systematic uncertainties of this approach come from the limited statistical power of 
the $Z\rightarrow \ell \ell$ recoil library, as is true with the other methods.

In this paper, we outline the implementation of this method. 
The method is tested using the $W$ boson mass and width measurements. 
Only the electron decay channel is discussed, but our method can 
also be used in the muon decay channel.
The detector and selection criteria are described in Section~\ref{sec-testbed}. 
The MC simulation samples used are described in 
Section~\ref{sec-samples}. We discuss the method in Section~\ref{sec-method}. 
In Sections~\ref{sec-rec-systs} and~\ref{sec-fullMC}, we assess the uncertainty on the $W$ boson 
mass and width measurements, and compare the performance of this method 
with that of a parameterized recoil model. The paper concludes in Section~\ref{sec-conclusion}. 

\section{The $W$ and $Z$ Boson Measurements}
\label{sec-testbed}
We evaluate the recoil library method by estimating biases and statistical and systematic 
uncertainties on the $W$ boson mass and width measurements in the electron channel.
The test is performed using simulations of the Run II D0 detector
at the Fermilab Tevatron, a \ppbar~collider with center of mass 
energy $\sqrt{s}=$1.96 TeV. Statistical uncertainties are estimated 
for a data sample corresponding to 1 fb$^{-1}$. 

\subsection{The D0 Detector}
The D0 detector~\cite{{Abachi:1993em},{bib:NIMRunII}} 
consists of a magnetic central-tracking system, comprised of a silicon 
microstrip tracker (SMT) and a central fiber tracker (CFT), both 
located within a 2~T superconducting solenoidal magnet. The SMT and CFT cover 
$|\eta_D| < 3.0$~\cite{bib:geo} and $|\eta_D| < 1.8$, respectively.

Three uranium/liquid argon calorimeters measure particle energies. The 
central calorimeter (CC) covers $|\eta_D|<1.1$, and two end calorimeters
(EC) extend coverage to $|\eta_D| \approx 4.2$. In addition to the preshower detectors, 
scintillators between the CC and EC cryostats provide sampling of developing showers 
at $1.1<|\eta_D|<1.4$. The CC is segmented in depth into eight layers. The first four 
layers are used primarily to measure the energies of photons and electrons and are 
collectively called the electromagnetic (EM) calorimeter. The remaining four layers, 
along with the first four, are used to measure the energies of hadrons. Most layers 
are segmented into $0.1 \times 0.1$ regions in $(\eta,\phi)$~\cite{bib:geo} space. 
The third layer of the EM calorimeter is segmented into $0.05\times0.05$ regions.

A muon system is located beyond the calorimetry and consists of a layer of tracking detectors
and scintillator trigger counters before 1.8 T iron toriods, followed by two similar layers 
after the toroids. Tracking at $|\eta_D|<1$ relies on 10 cm wide drift tubes, while 1 cm 
mini-drift tubes are used at $1<|\eta_D|<2$.

Scintillation counters covering $2.7<|\eta_D|<4.4$ are used to measure luminosity 
and to indicate the presence of an inelastic collision in beams crossing.

\subsection{Measurement strategies for $M_W$ and $\Gamma_W$}
\label{sec-fit}
The $W$ boson mass is measured from distributions of the following 
observables: the electron transverse momentum $\vec{p}_T^{~e}$, the 
missing transverse energy $\vmet$, and the transverse mass $M_T$, given by
\begin{equation}
   M_T = \sqrt{2|\vec{p}_T^{~e}|~|\vmet| [1-\cos(\Delta \phi)]}, 
\label{eq:mT}
\end{equation}
where $\Delta \phi$ is the opening angle between $\vec{p}_T^{~e}$ and $\vmet$ in the transverse plane. 
The data distributions are compared with probability density functions 
from MC simulations generated with various input $W$ boson mass values (``templates''). 
A binned negative log-likelihood method is used to extract $M_W$. The $W$ boson 
width is measured using a similar method, except that only events in the 
high tail region of the $M_T$ distribution are used.
For the mass measurement, the fit ranges we used are [30, 48] GeV for the 
$|\vec{p}_T^{~e}|$ and $|\vmet|$ distributions, and [60, 90] GeV for 
the $M_T$ distribution. For the width measurement, we fit 
the $M_T$ distribution over the range [100, 200] GeV.

\subsection{Selection criteria}
\label{subsec-select}
A $W$ boson candidate is identified as an isolated electromagnetic cluster accompanied 
by large $|\vmet|$. The electron candidate is required to have a
shower shape consistent with that of an electron, $|\vec{p}_T^{~e}| > 25$ GeV, and 
$|\eta_{D}| < 1.05$. To further reduce backgrounds, the electron candidate is required to be 
spatially matched to a reconstructed track in the central tracking system. 
Additionally, we require $|\vmet|>25$ GeV, $|\vec{u}_T|<15$ GeV, and $50<M_T<200$ GeV. 
$Z$ boson candidates are identified as events containing two such electrons with di-electron invariant 
mass $70<M_{ee}<110$ GeV and $|\vec{u}_T|<15$ GeV. 
The selection on $|\vec{u}_T|$ helps to suppress background and to 
reduce the sensitivity of the measurement to uncertainties on the 
detector model and the theoretical description of the $p_T^W$ distribution. Since 
the $Z$ sample has fewer events at high $p_T^Z$, the detector and 
theoretical models are best constrained at low boson $p_T$. 

For this analysis, both electrons from the $Z$ boson are required to be in 
the central region of the calorimeter because the unfolding requires 
well-understood detector resolutions.

\section{MC Simulation Samples}
\label{sec-samples}
In this paper we use three different MC simulations. 
Two of these are fast MC simulations and the third is a 
detailed full MC simulation using {\sc geant}~\cite{bib:Geant}. 
The two fast MC simulations are built around a common event generator 
and parametric model for the electron measurement, but with different 
recoil models. One uses a traditional parameterized method to model
the recoil system, which we call ``the parameterized recoil method". The 
other uses our new method, which we call ``the recoil library method".
The full MC $\zee$ sample has the equivalent of 6.0 fb$^{-1}$ in 
integrated luminosity, and the full MC $\wen$ sample corresponds to 2.5 fb$^{-1}$.

For both fast MCs, the {\sc pythia}~\cite{bib:Pythia} event generator is used to simulate 
the production and decay of the $W$ boson, as well as any final state radiation (FSR) photons. 
FSR photons, if sufficiently close to the electron,  
are merged with the electron. After the event kinematics are generated at the 
four-vector level, detector efficiencies and energy response and resolution for the electron 
are applied. These parameterizations are measured using $\zee$ events 
from either collider data or full MC, depending on the study. 
A parametric energy dependent model for resolution effects is used. 
Parameterized efficiencies for data selection are prepared for comparing with 
either data or full MC as a function of electron $|\vec{p}_T^{~e}|$, $\eta^e$, 
the component of the recoil along the electron direction, the total hadronic 
activity in the event, and the reconstructed $z$ coordinate along the beam line where the
hard scattering occurred. The recoil system is then modeled either 
using the recoil library or the parameterized model.

The parameterized recoil method models the detector response to the hard 
recoil using a two-dimensional parameterization of the response 
(both magnitude and direction) as estimated using {\sc geant}-simulated 
$Z \rightarrow \nu \bar{\nu}$ events.
The underlying event is modeled using the measured $~\vmet$ distribution from 
data taken with a trigger that requires energy in the luminosity monitors 
(``minimum bias events''), and pileup and additional interactions are modeled using 
the measured $\vmet$ distributions from unsuppressed data taken on 
random beam crossings (``zero bias events''). These are combined with the hard recoil, and data-tuned 
corrections are applied to account on average for correlations between the ``hard'' and ``soft'' recoil. 
The correction parameters are tuned to $\zee$ control samples. The parametric methods 
of modeling the recoil are further discussed in Refs.~\cite{{bib:RunIImass}} 
and resemble approaches used in earlier D0 and CDF measurements at the 
Tevatron~\cite{{cdfw},{d0w},{cdfw2}}. The recoil library method of modeling the recoil is discussed 
in detail in Section~\ref{sec-method}.

The {\sc geant}-based MC simulation also uses {\sc pythia} to simulate the production and 
decay of the $W$ boson, as well as the underlying event and any FSR photons. 
These events are then propagated through a detailed description of the detector. 
Zero bias collider data collected by the D0 detector
with a similar instantaneous luminosity profile as the $\wen$ collider 
data sample are overlaid on the full MC simulation to model additional 
collisions and noise in the detector. These events are processed through 
the same full set of D0 reconstruction programs as data.

\section{The Recoil Library Method}
\label{sec-method}

\subsection{Overview}
The recoil library is built from $\zee$ events. 
Because the electron energies and angles are well measured, 
the measured $\vec{p}_T^{~Z}$ from the two electrons provides a good first 
approximation of the true $\vec{p}_T^{~Z}$. An unfolding procedure allows 
the transformation of the two-dimensional distribution of the measured 
$|\vec{p}_T^{~Z}|$ and measured $|\vec{u}_T|$ to that of the true $|\vec{p}_T^{~Z}|$ 
and measured $|\vec{u}_T|$.
The opening angle between the measured $\vec{p}_T^{~Z}$ and the measured $\vec{u}_T$ 
is also unfolded to the opening angle between the true $\vec{p}_T^{~Z}$ and the 
measured $\vec{u}_T$ during this procedure. A map between the true $|\vec{p}_T^{~Z}|$, 
the measured $|\vec{u}_T|$, and the scalar $E_T$ ($SE_T$), which is defined as the scalar sum
of the transverse energies of all calorimeter cells except those that belong to 
the reconstructed electrons, is also produced. This map is not used by the recoil model, but
is needed by the electron efficiency model. The final result of the recoil library is 
the $\vec{u}_T$ for an event, referenced to the true $\vec{p}_T^{~Z}$. This vector 
substitutes for the equivalent vector obtained in the parametrized recoil model. All 
further corrections for efficiencies due to the recoil system are the same for 
both the recoil library and the parametrized recoil model.

\begin{figure*}[htbp]
  \centering
    \includegraphics[scale=0.4]{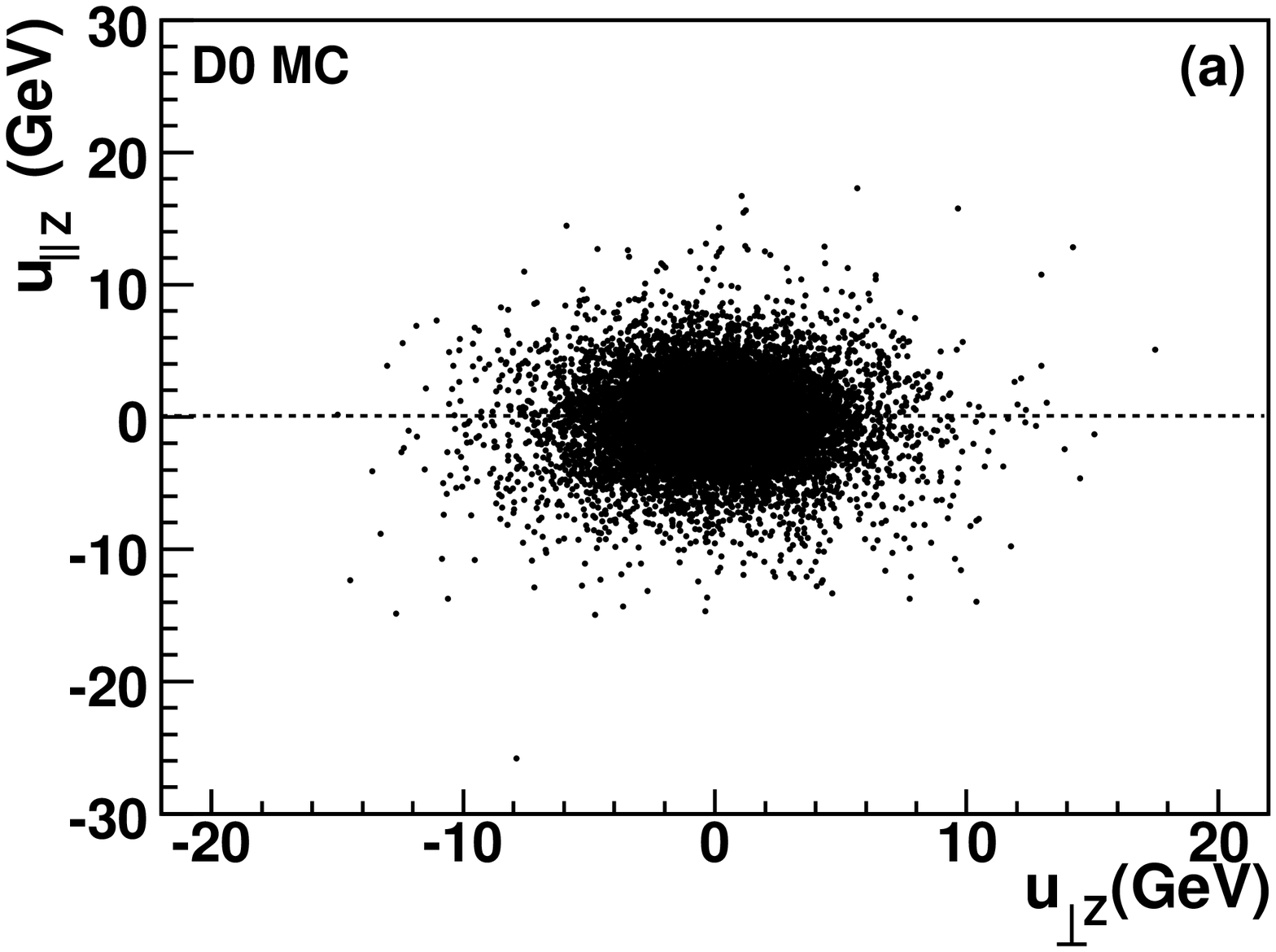}\hfil
    \includegraphics[scale=0.4]{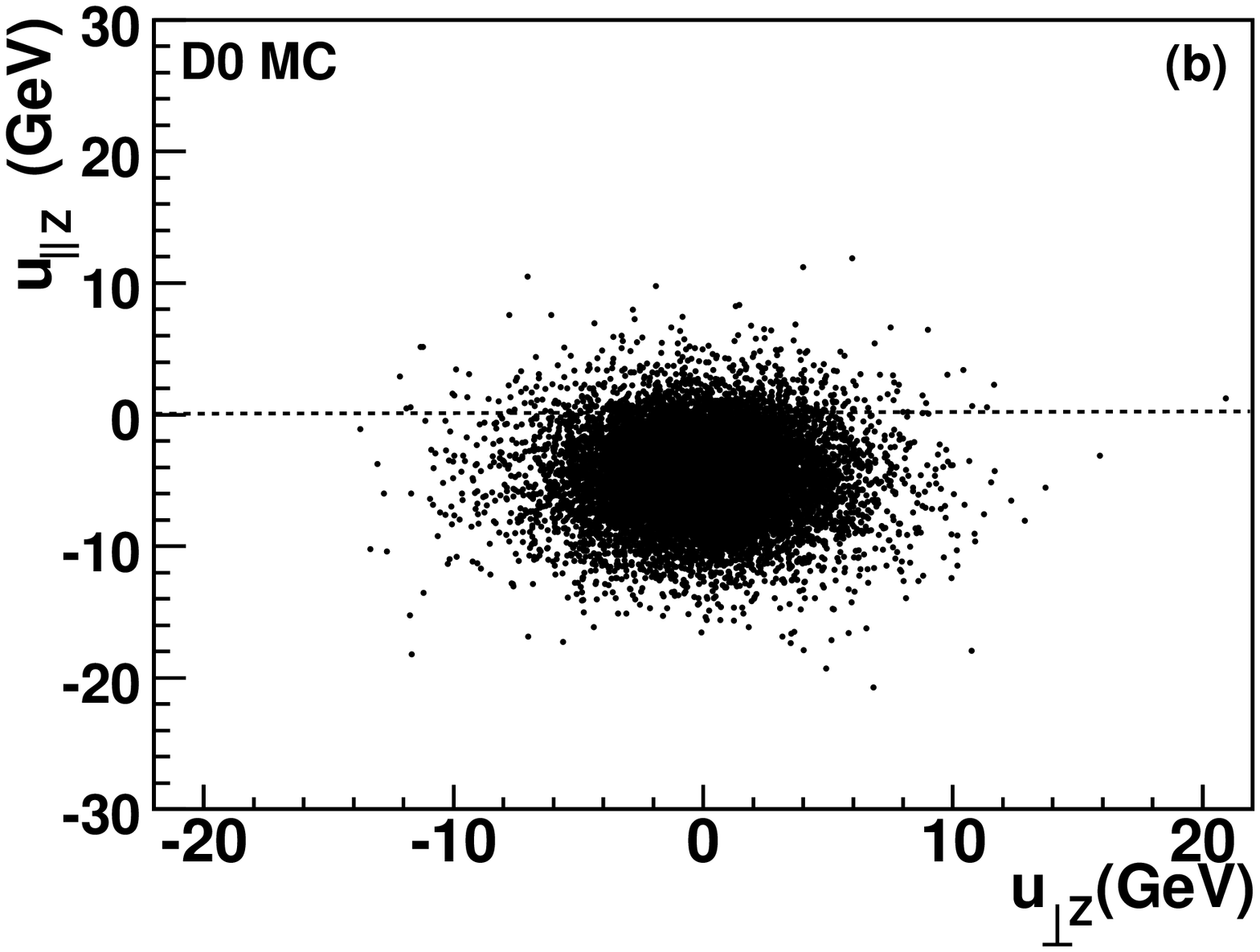}\hfil    
    \includegraphics[scale=0.4]{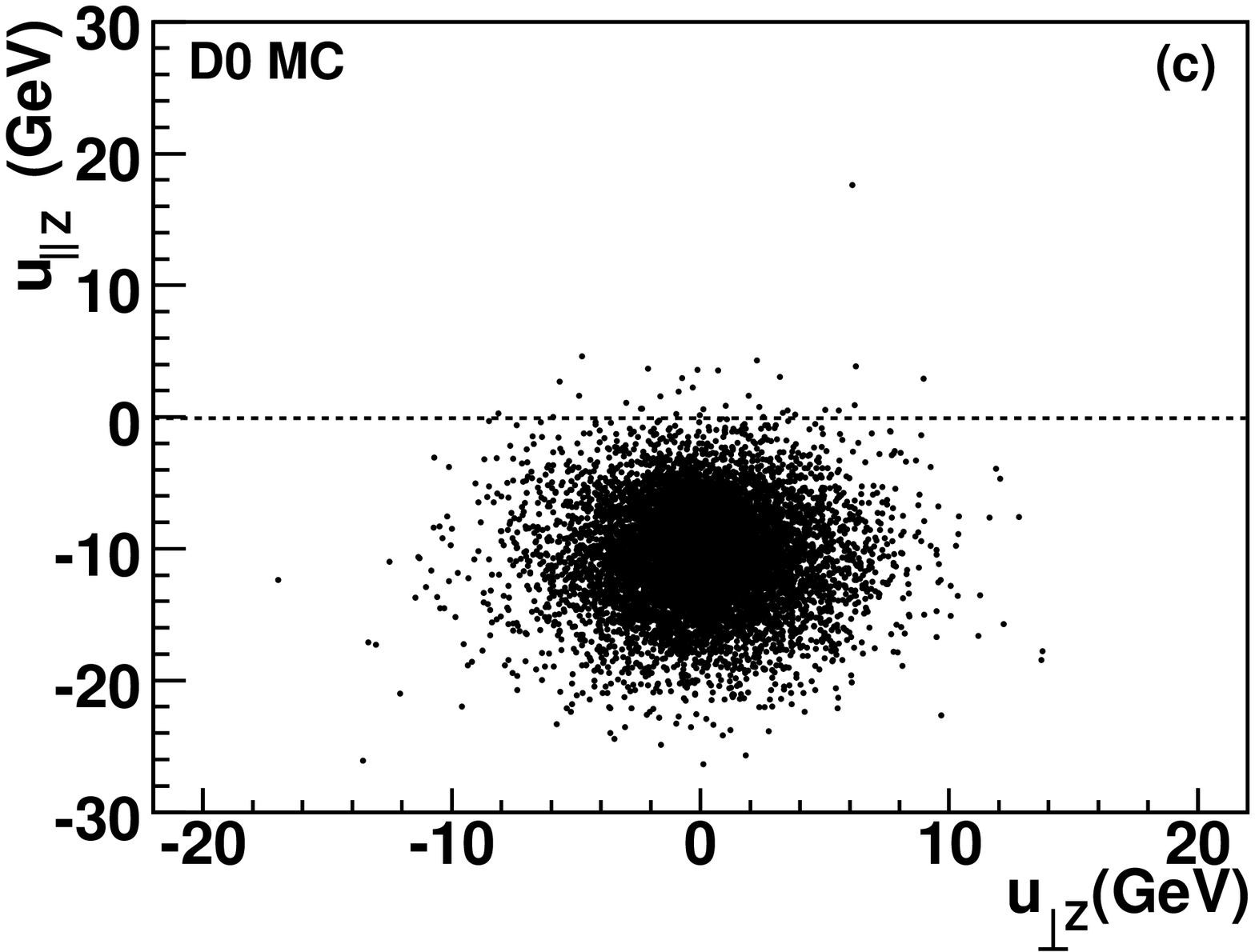}
    \caption{Examples of the distribution of the component of the measured recoil parallel ($u_{\parallel Z}$) and 
perpendicular ($u_{\perp Z}$) to the $Z$ boson direction for three different bins in true $|\vec{p}_T^{~Z}|$ 
(centered at (a) 0.4, (b) 10, and (c) 29 GeV). Each dot represents $\vec{u}_T$ from a single event in the library.}
  \label{fig:examplemap}
\end{figure*}

Figure~\ref{fig:examplemap} shows some examples of the distribution of the component of the measured recoil 
along the $Z$ boson direction and perpendicular to the $Z$ boson direction.

\subsection{Preparing the recoil library}
Before producing a binned recoil library, certain event-by-event 
corrections must be applied to the measured recoil system. We need 
to remove any biases in the measured recoil distribution due to the 
$Z$ boson selection requirements. Electron identification requirements, 
for example, preferentially reject events with significant hadronic activity.
Events with significant hadronic activity also have poorer recoil resolution
than events with little hadronic activity. 
Since $Z$ bosons contain two electrons 
while $W$ bosons only have one, the bias will not be the same.
The electrons from $Z$ boson decays
also have a higher average $|\vec{p}_T^{~e}|$ and a different $\eta^e$ 
distribution than electrons from $W$ boson decays. 
To account for this, we remove the biases from 
the $Z$ boson selection, and then, when a $W$ candidate is made using the recoil 
library, the biases appropriate for a $W$ candidate are added, as described in 
Ref.~\cite{bib:RunIImass}. In this section, we describe these corrections to 
the $Z$ boson sample.

\label{sec-uparcorr}
\subsubsection{Removing the two electrons from $Z$ boson events}
The recoil system for $\zee$ events is defined as the energies in all calorimeter cells
excluding those that belong to the two electrons.  
Since the recoil system will in general deposit energy in these cells, excluding 
them biases the component of the recoil along the electron's direction. We correct this effect 
by adding back an approximation of the underlying energy.

This correction (denoted by $\Delta u_{\parallel}^e$) 
depends on $u_{\parallel}^e$ (the projection of $\vec{u}_T$ along the 
electron transverse direction), instantaneous luminosity, and electron 
$\eta^e$, and is estimated using the energies deposited in equivalent 
cells that are separated in $\phi$ from the electron in $\wen$ events.
In addition to correcting for the recoil energy under the 
electron cluster, we also correct for electron energy that leaks 
out of the cluster. For $Z$ boson events, these corrections are made for both electrons. 
In Section~\ref{sec-rec-systs} we estimate the uncertainty due to these corrections. 

\subsubsection{Minimizing the effects of FSR photons}
\label{subsubsec-fsrmin}
The full MC simulation indicates that roughly 6\% of the $\zee$ events contain FSR photons 
with $E_T^\gamma > 400$ MeV that are sufficiently far from the electrons that the 
electron clustering algorithm at D0 does not merge them with a 
reconstructed electron. These photons are thus incorrectly included in the 
measurement of $\vec{u}_T$, instead of in $\vec{p}_T^{~Z}$, resulting in a correlated bias.
Since $\zee$ events contain more FSR photons than $\wen$ events do, 
the recoil library built using $Z$ bosons will contain on average 
larger contributions from FSR photons. 

Ideally, these FSR photons could be removed
from the recoil file, and the effect could be separately
modeled within the fast MC simulation. Since it is difficult to 
identify these FSR photons on an event-by-event basis, 
the effect is reduced by raising the lower limit on the effective
reconstructed di-electron invariant mass to 85 GeV, reducing the fraction 
of events with a high $E_T$ FSR photon by 25\%. 

The effect of the remaining photons is small because, for a low $p_T$ $W$ boson,
$M_T \approx 2 |\vec{p}_T^{~e}|+ u_{\parallel}^e$.
Therefore, the photons will create a bias on the mass only if they  produce
a bias in the component of $\vec{u}_T$ parallel to the electron direction. 
While the overlaid recoil is rotated so that the direction of its corresponding
$Z$ boson matches that of the simulated $W$ boson, the directions of the 
decay electrons from $Z$ and $W$ are largely uncorrelated, and the bias
is mostly canceled for measurements using the $M_T$ spectrum.
In Section~\ref{sec-rec-systs} the bias due to the FSR photons is estimated. 

\subsubsection{Correcting for electron selection efficiencies}
\label{subsec-efficiences}
The selection criteria for $W$ and $Z$ candidates can introduce biases between 
the electron and the recoil system. Since the kinematic and geometric properties 
of $W$ candidates are not identical to those of $Z$ candidates, they have different biases. 

The two components of the electron selection efficiency model that 
most strongly affect these biases are the $SE_T$ efficiency and the $u_{||}^e$  efficiency. 
The $SE_T$ efficiency describes the electron identification probability as 
a function of the overall activity in the detector. 
The $u^e_{\parallel}$ efficiency describes the probability of electron 
identification as a function of $u^e_{\parallel}$. This probability 
decreases with increasing hadronic activity along the electron direction. 

Since the recoil library is built from $\zee$ events, we need to remove the 
biases introduced by the selection requirements on the two electrons.
We correct for the efficiencies by weighting each 
event in the $Z$ boson recoil library by 
$1/ \epsilon_{u_{\parallel}^e} \times 1/\epsilon_{SE_T}$, 
where $\epsilon_{u^e_{\parallel}}$ 
is the product of the $u^e_{\parallel}$ efficiencies and 
$\epsilon_{SE_T}$ is the product of the $SE_T$ efficiencies 
for the two electrons in each $Z$ candidate.

When $W$ boson events are produced from a fast MC using the recoil library, 
the map between the true $\vec{p}_T^{~Z}$, measured $\vec{u}_T$, and $SE_T$ is used to 
introduce the biases appropriate for $W$ bosons from these efficiencies. To simulate 
a $W$ boson event, a random recoil is chosen from the recoil library corresponding 
to the true $W$ boson $p_T^W$, and a random $SE_T$ is chosen 
from the $SE_T$ distribution corresponding to the true $W$ boson $p_T^W$ 
and the chosen recoil $\vec{u}_T$. The $u^e_{\parallel}$ efficiency and $SE_T$ 
efficiency are then applied to the electron from $W$ boson decays.

\subsection{Unfolding method}
\label{sec-unfold}
After the recoils have been corrected as discussed above, the transformation
from measured $\vec{p}_T^{~Z}$ and measured $\vec{u}_T$ to true $\vec{p}_T^{~Z}$ 
and measured $\vec{u}_T$ is done using a Bayesian unfolding technique. 

\subsubsection{Multidimensional unfolding using Bayes's Theorem}
Unfolding is a mathematically challenging problem, since 
it involves the reversal of a random process. Because a given true state 
can migrate to many measured states and many different true states can 
migrate to the same measured state, we cannot unfold detector effects 
on an event-by-event basis. Rather, unfolding methods typically work with binned distributions.

For the recoil library method, we chose to use a Bayesian unfolding 
approach~\cite{bib:bayes}. This approach suits our needs because it is 
intuitive, simple to implement, and easy to extend to the multidimensional case. The Bayesian 
technique uses conditional probabilities to determine the probability 
that a given measured state corresponds to a particular true state.

Consider a distribution of initial states $I_i$, 
$\lbrace{i=1,2,...,N_I \rbrace}$, given by $P(I_i)$ (the probability of events
with initial state $I_i$) and a distribution of final 
states $F_j$, $\lbrace{j=1,2,...,N_F \rbrace}$, 
given by $P(F_j)$ (the probability of events 
with final state $F_j$), given the measured distribution 
$P(F_j)$, and the probability for each initial state to migrate 
to each final state $P(F_j|I_i)$, we can determine the distribution of initial states $P(I_i)$ using
\begin{equation}
P(I_i) = \sum_{j=1}^{N_F}P(I_i|F_j) P(F_j).
\label{eq:init_dist}
\end{equation}

Using simulations, we can calculate $P(I_i|F_j)$ from
$P(F_j|I_i)$, the likelihood of a true state 
fluctuating to a measured state, using Bayes's theorem, which is
\begin{equation}
P(A|B) =  \frac{P(B|A)P(A)}{P(B)}.
\label{eq:bayes}
\end{equation}
For our particular example, with $N_I$ initial states and $N_F$ final states, Bayes's theorem gives us
\begin{equation}
P(I_i|F_j) =  \frac{P(F_j|I_i)P(I_i)}{ \sum_{k=1}^{N_I} P(F_j|I_k)P(I_k)}.
\label{eq:bayes_manystates}
\end{equation}

We can interpret this equation as follows: the probability that a given final state $F_j$ comes from a 
particular initial state $I_i$ is proportional to the probability density of state $I_i$ multiplied 
by the probability that $I_i$ migrates 
to $F_j$. The denominator normalizes the distribution.

Our Bayesian method requires us to make assumptions regarding the distribution of initial 
states, $P(I_i)$. Although we only use $P(I_i)$ to calculate the weights used for the 
measured data, the quality of the unfolding could depend on $P(I_i)$. To minimize 
this effect, the method is applied iteratively, starting with a reasonable prior 
for the distribution with $P_0(I_i)$, and with each successive iteration using 
the previous iteration's unfolded distribution as the new input. 
As a cross-check, the method is applied with several different initial 
$P_0(I_i)$ distributions. The iteration procedure is:
\begin{enumerate}
\item Choose an initial seed input distribution for $P_0(I_i)$.
\item Using $P_0(I_i)$ and $P(F_j|I_i)$, compute the weights $P(I_i|F_j)$, as 
derived using the Bayesian equation shown in Eq.~\ref{eq:bayes_manystates}.
\item Using these weights, recalculate the unfolded distribution $P_1(I_i)$ from the 
relationship $P_1(I_i) = \sum_{j=1}^{N_F} P_0(F_j)P(I_i|F_j)$ described in Eq.~\ref{eq:init_dist}.
\item Repeat the above steps with $P_1(I_i)$ as the starting distribution.
\item Iterate until the unfolded $P(I_i)$ converges.
\end{enumerate}  

\subsubsection{Unfolding the recoil distribution}
\label{subsubsec-recunfold}

For our application, the recoil vector is described by the coordinates $(|\vec{u}_T|,\psi^t)$, 
where $|\vec{u}_T|$ is the magnitude of the measured recoil transverse momentum, and 
$\psi^t$ is the opening angle between the measured recoil and the true boson 
direction in the transverse plane. These recoil vectors are stored 
in an array of two-dimensional recoil  histograms (binned in $|\vec{u}_T|$ and
$\psi^t$). Each histogram corresponds to a discrete bin in true $|\vec{p}_T^{~Z}|$
with bins of 0.25 GeV for small $|\vec{p}^{~Z}_T|$ ($|\vec{p}^{~Z}_T| <50$ GeV)
and larger bins at larger $|\vec{p}^{~Z}_T|$. 

In the implementation of Eq.~\ref{eq:bayes_manystates}, the initial state $I$ is
specified by $[(|\vec{p}^{~Z}_T|)^t_i,\psi^t_j, {(|\vec{u}_T|)}_k]$ and the final state $F$ is given by 
$[(|\vec{p}^{~Z}_T|)^s_m, \psi^s_n, {(|\vec{u}_T|)}_k]$, where $(|\vec{p}^{~Z}_T|)^t$ 
is the true $Z$ boson transverse momentum, $(|\vec{p}^{~Z}_T|)^s$ is the smeared $Z$ boson transverse momentum, 
and ${\psi}^s$ is the opening angle between the measured recoil and the smeared $Z$ boson direction 
in the transverse plane.

\begin{figure*}[htbp]
\centering
\includegraphics [scale=0.4] {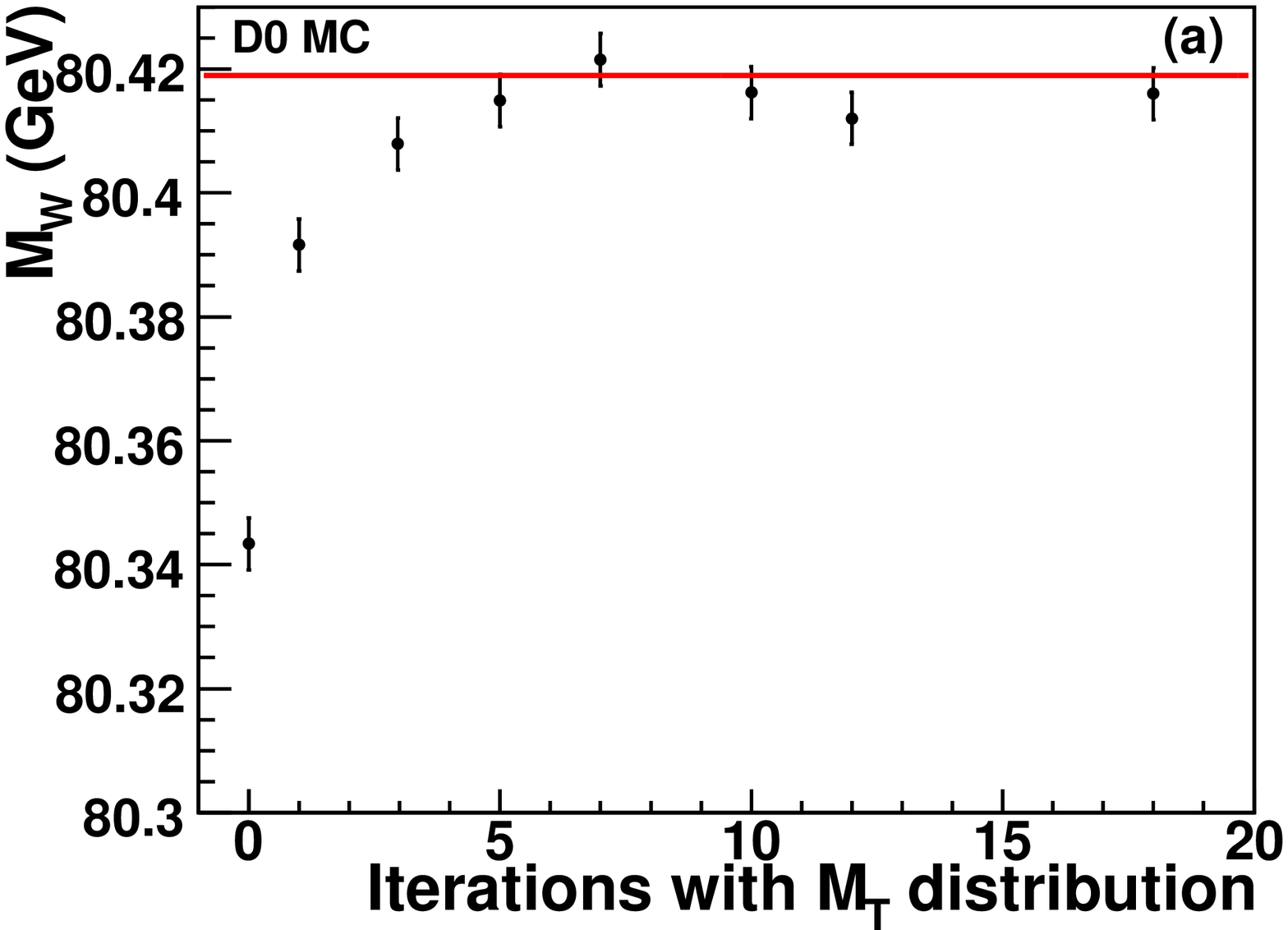} 
\includegraphics [scale=0.4] {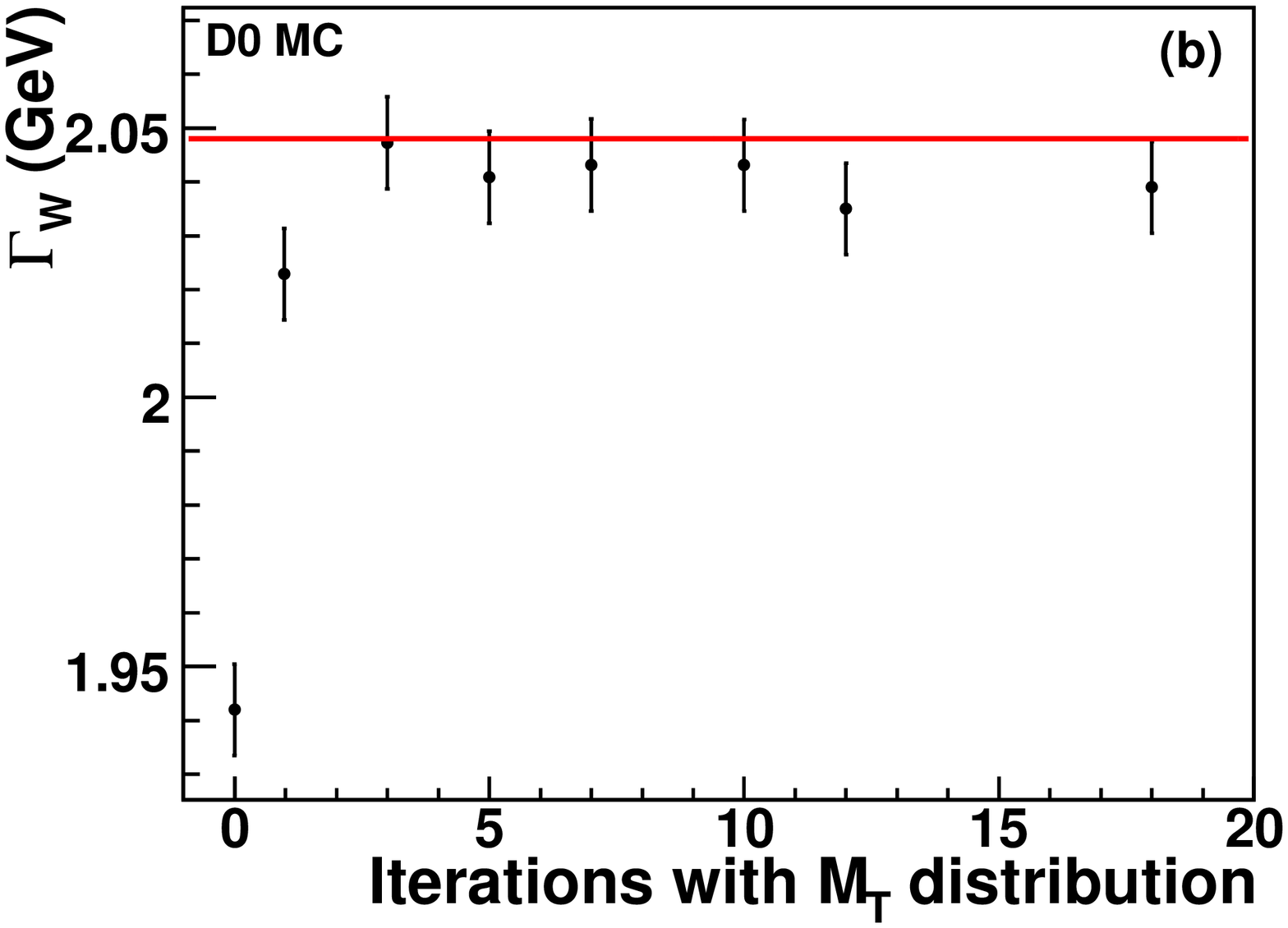} 
\caption{Estimated (a) $W$ boson mass and (b) $W$ boson width in fast MC 
using the $M_T$ distribution, as a function of number of iterations used in the 
unfolding. The red line indicates the input values of $W$ boson mass and width 
in the fast MC. The default number of iterations used is 10.}
\label{fig:converge}
\end{figure*}

We start with an initial seed distribution that is flat in 
$(|\vec{p}^{~Z}_T|)^t$, $\psi^t$, and $|\vec{u}_T|$. We find that it takes fewer 
than 10 iterations for the unfolding method to converge. 
Figure~\ref{fig:converge} shows the convergence of the $W$ boson mass and width 
obtained from the $M_T$ distribution, as a function of iteration number in fast 
MC studies. The final value achieved agrees well with the input value.
The systematic uncertainty on the $W$ boson mass and width due to the unfolding 
procedure is discussed further in Section~\ref{sec-rec-systs}.

\begin{figure}[htbp]
  \begin{center}
    \includegraphics[scale=0.5]{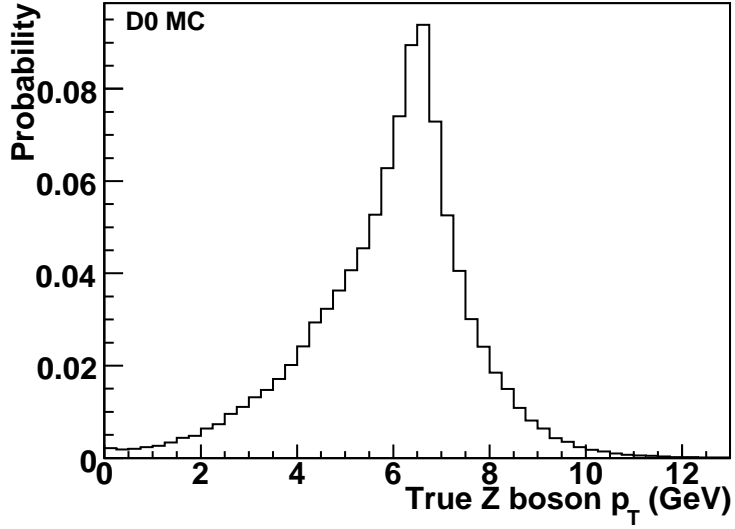}
  \end{center}
  \caption{The distribution of the probabilities that a reconstructed $|\vec{p}^{~Z}_T|$ of 7 GeV with 
corresponding $|\vec{u}_T|$ of 3.5 GeV comes from various true $|\vec{p}^{~Z}_T|$ bins.}
  \label{fig:prob_dist}
\end{figure}

\begin{figure*}[htbp]
  \centering
    \includegraphics[scale=0.41]{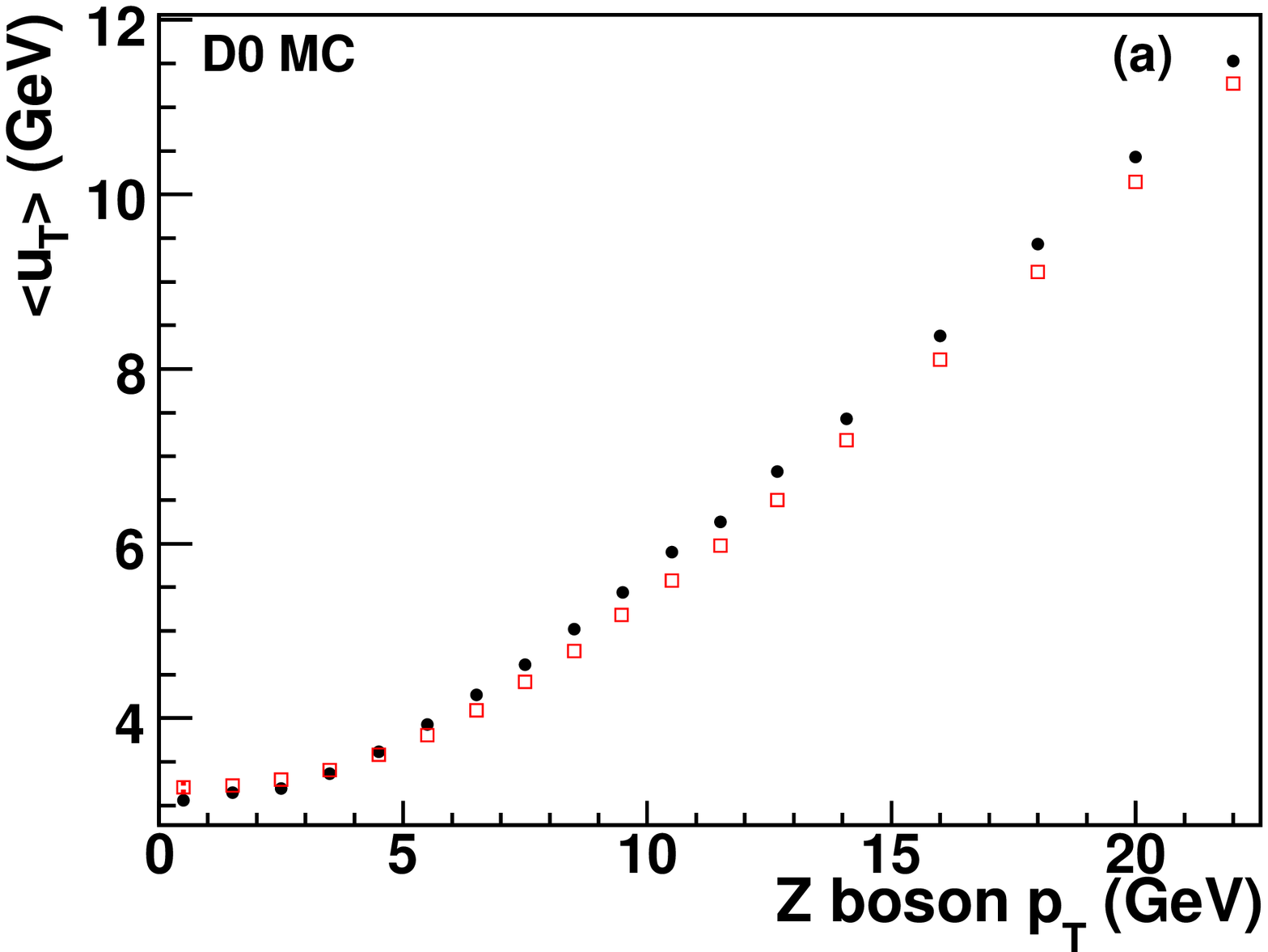}\hfil
    \includegraphics[scale=0.41]{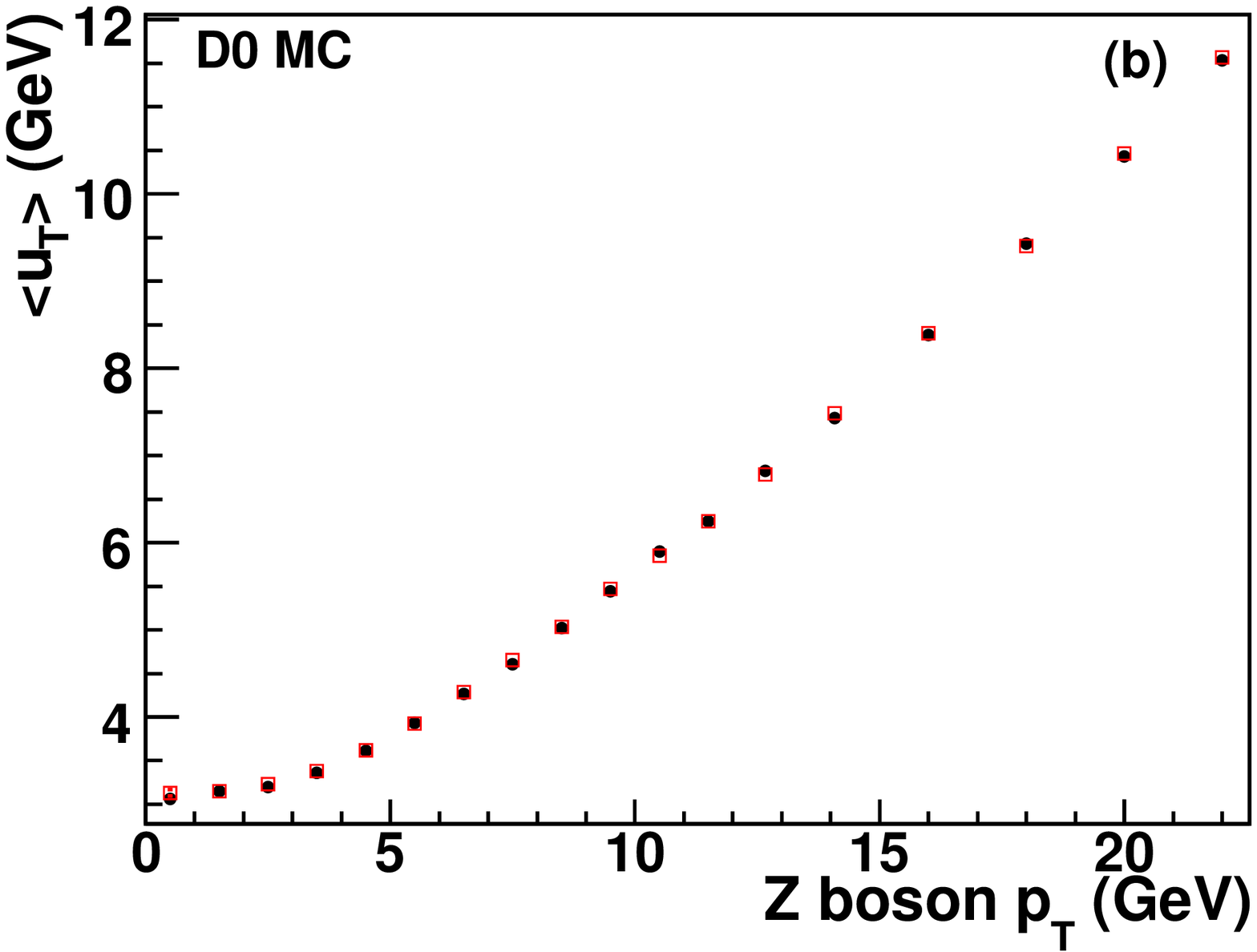}
  \caption{Mean recoil $|\vec{u}_T|$ 
versus true $|\vec{p}^{~Z}_T|$ (black filled points) and 
mean recoil $|\vec{u}_T|$ versus the estimate of the true $\vec{p}_T^{~Z}$ using the two electrons (red open boxes) when using (a) the two smeared electrons directly and (b) the unfolded map.}
  \label{fig:RpT_unfolded}
\end{figure*}

\begin{figure*}[htbp]
  \centering
    \includegraphics[scale=0.41]{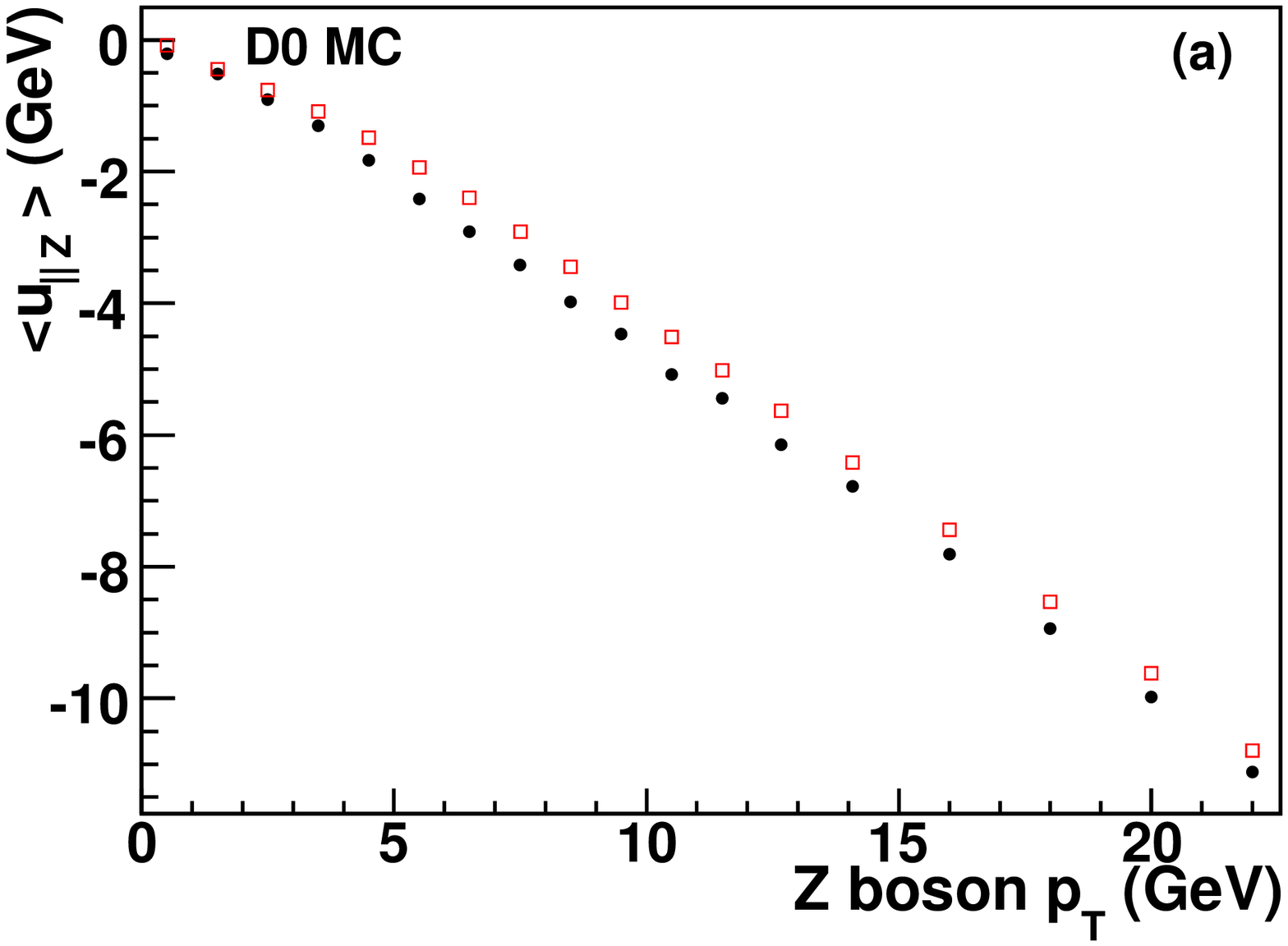}\hfil
    \includegraphics[scale=0.41]{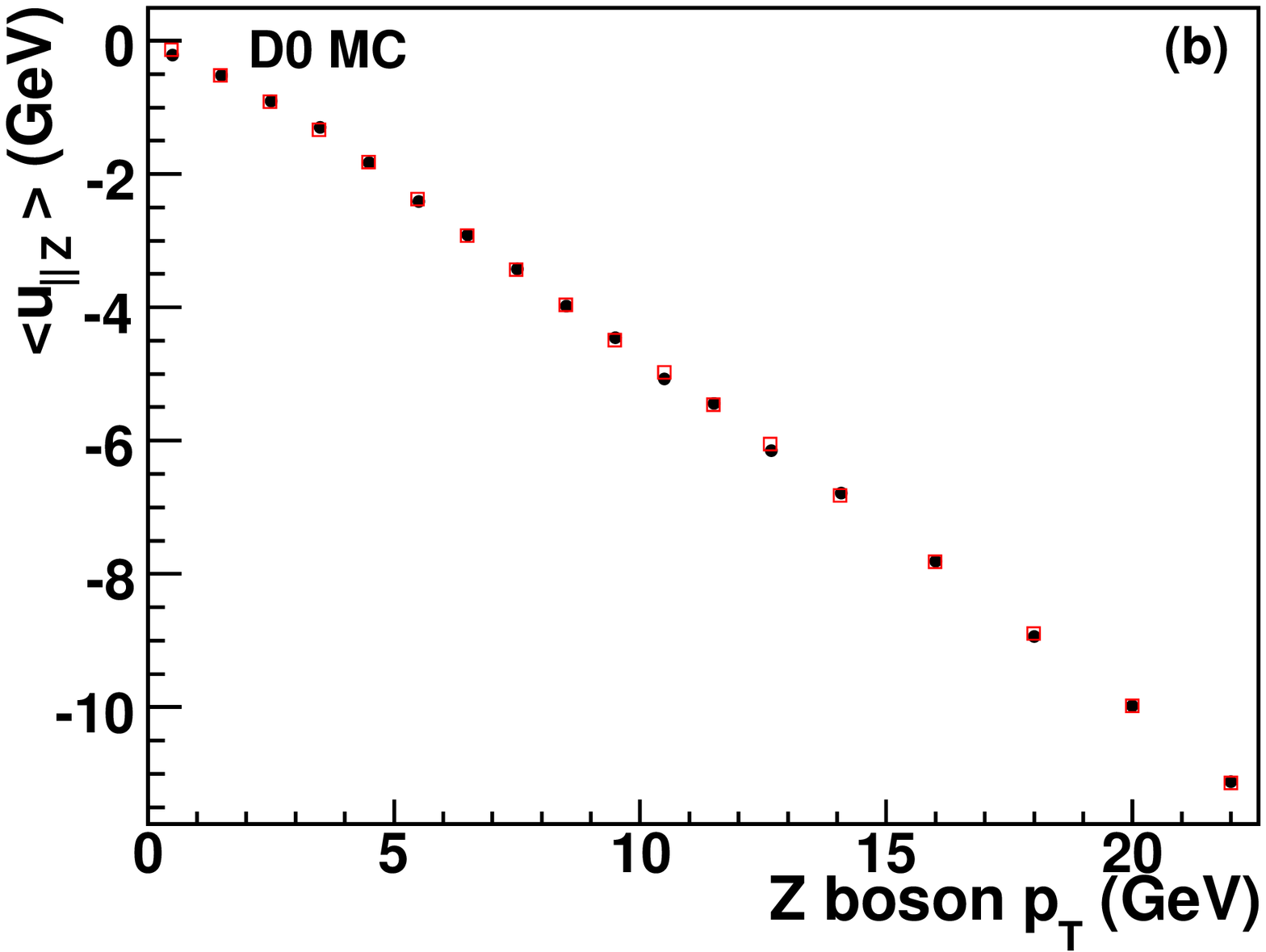}
  \caption{Mean projection of the recoil along the $Z$ 
boson direction ($<u_{\parallel Z}>$) versus true $|\vec{p}^{~Z}_T|$ (black filled points) and mean projection of the recoil along the boson direction versus the estimate of the true $\vec{p}_T^{~Z}$ using the two electrons (red open boxes) when using (a) the two smeared electrons directly and (b) the unfolded map.}
  \label{fig:RpY_unfolded}
\end{figure*}

\begin{figure*}[htbp]
  \centering
    \includegraphics[scale=0.41]{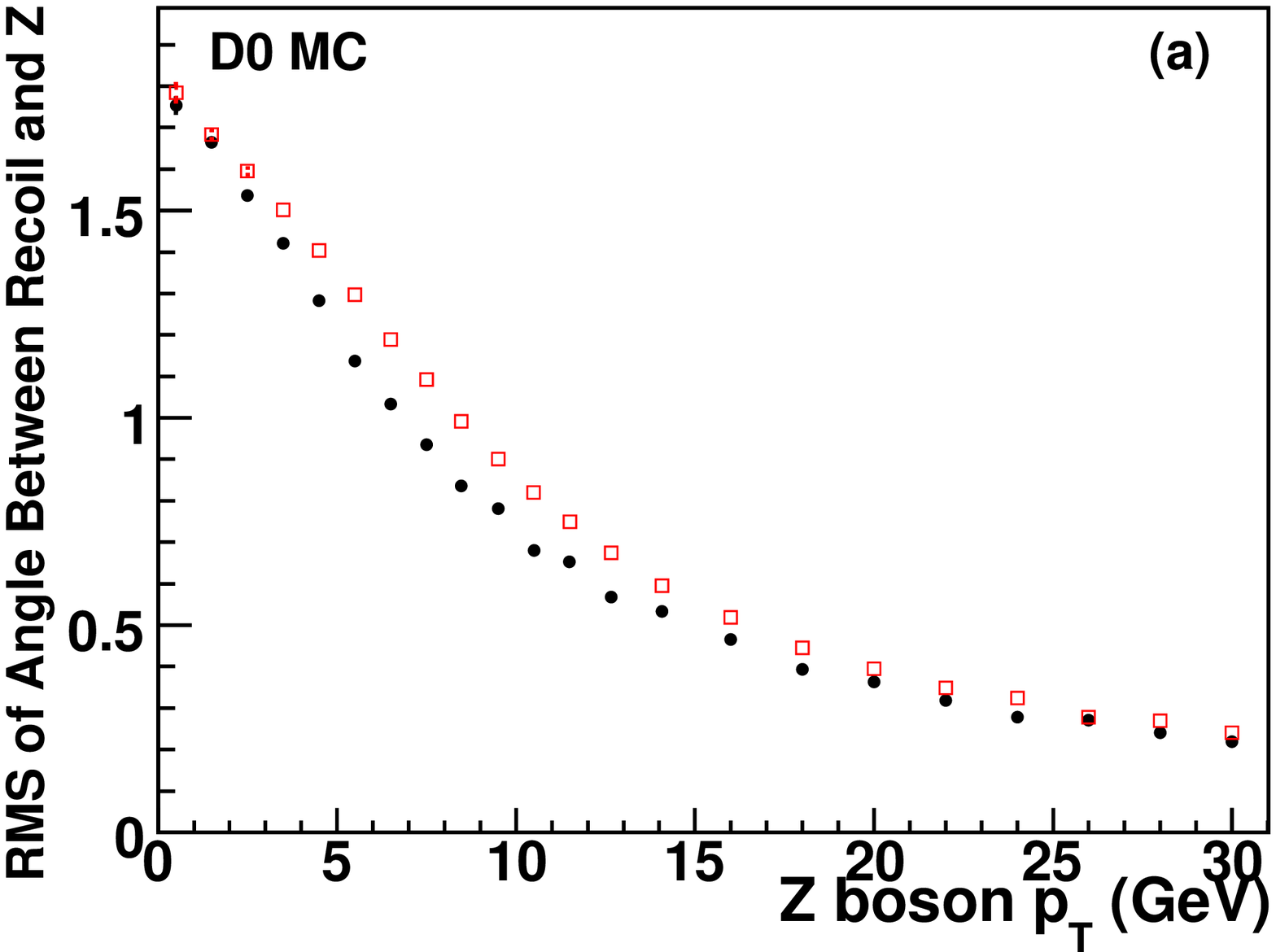}\hfil
    \includegraphics[scale=0.41]{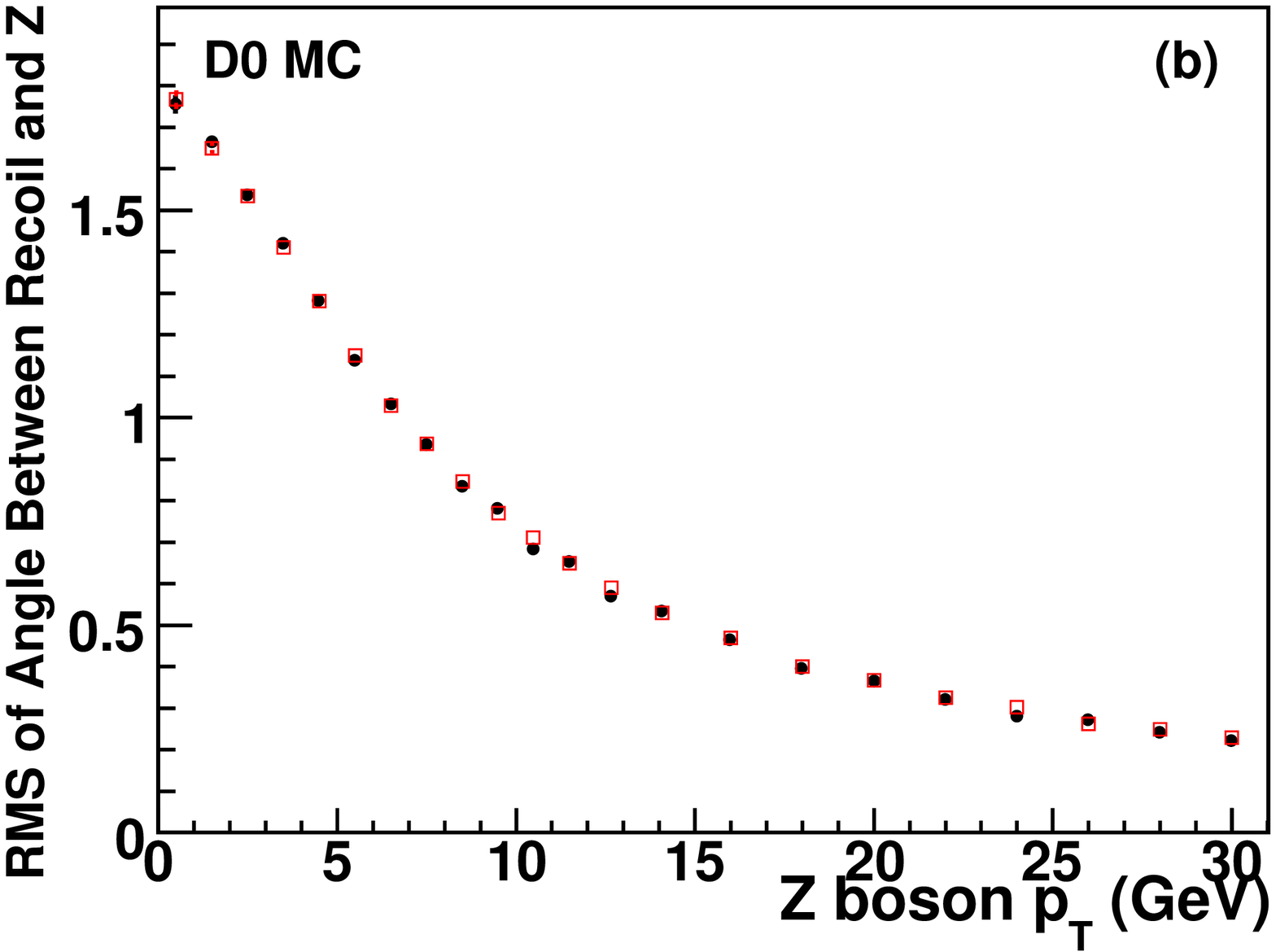}
  \caption{RMS of the opening angle between $\vec{u}_T$ 
and $\vec{p}^{~Z}_T$ versus true $\vec{p}^{~Z}_T$ (black filled points) and RMS of the opening angle between the 
   recoil and the boson versus the estimate of the true $|\vec{p}^{~Z}_T|$ using the two electrons (red open boxes) when using (a) the two smeared electrons directly and (b) the unfolded map.}
  \label{fig:RphiRMS_unfolded}
\end{figure*}

\begin{figure*}[htbp]
  \centering
    \includegraphics[scale=0.41]{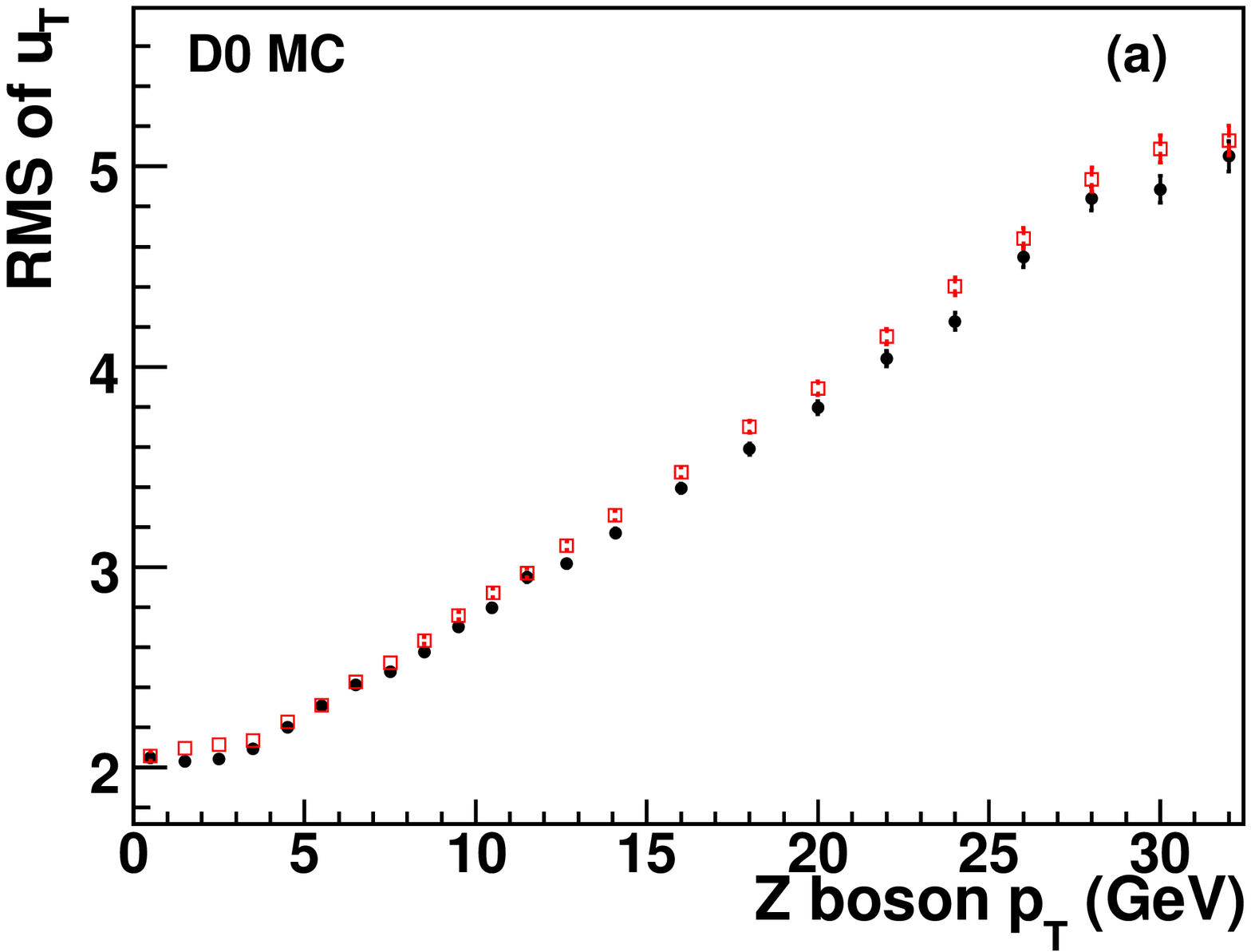}\hfil
    \includegraphics[scale=0.41]{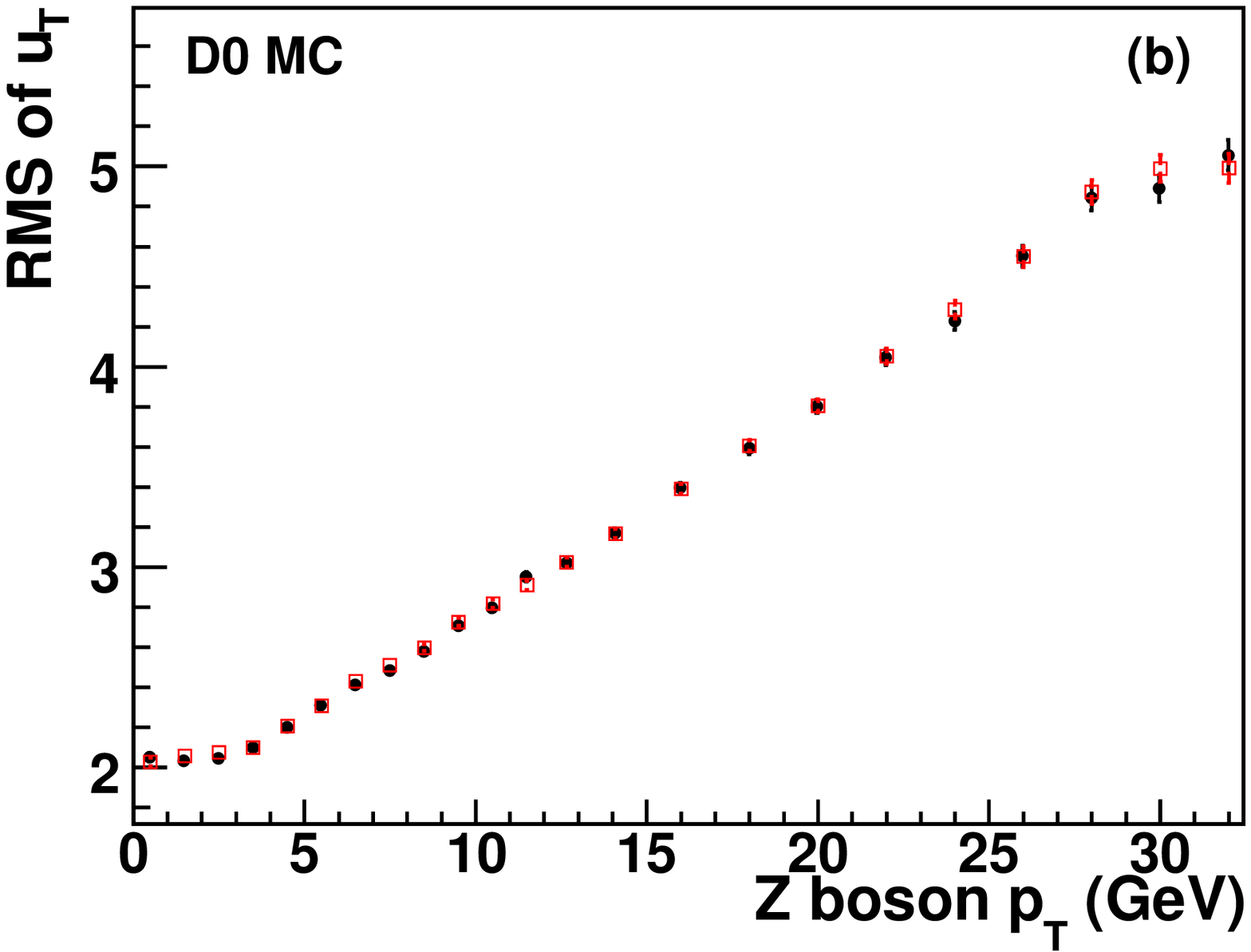}
  \caption{RMS of the recoil $|\vec{u}_T|$ versus true $|\vec{p}^{~Z}_T|$ (black filled points) and RMS of the recoil $|\vec{u}_T|$ versus the estimate of the true $|\vec{p}_T^{~Z}|$ using the two electrons (red open boxes) 
   when using (a) the two smeared electrons directly and (b) the unfolded map.}
  \label{fig:RpTRMS_unfolded}
\end{figure*}

\begin{figure*}[htbp]
  \centering
    \includegraphics[scale=0.38]{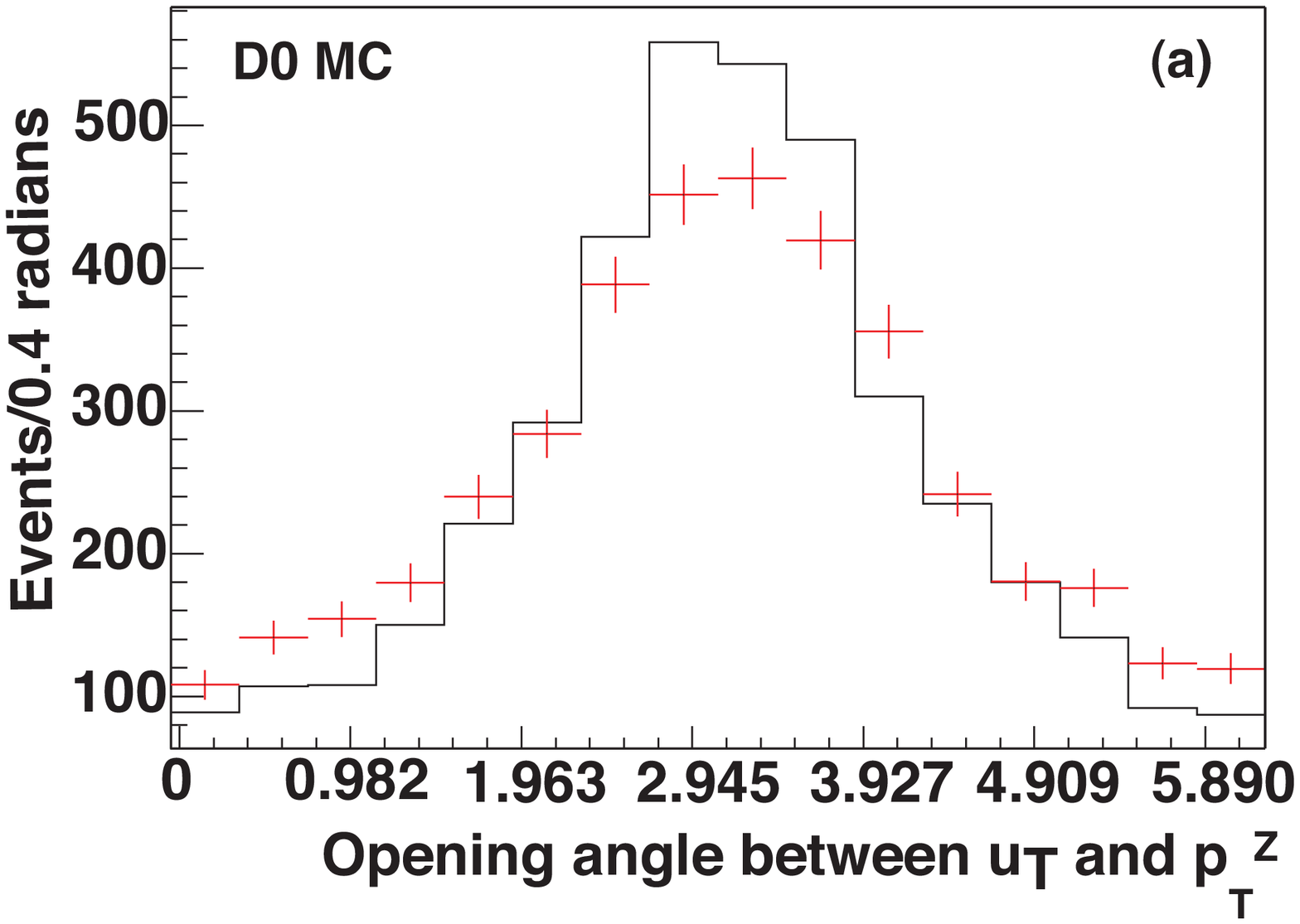}
    \includegraphics[scale=0.38]{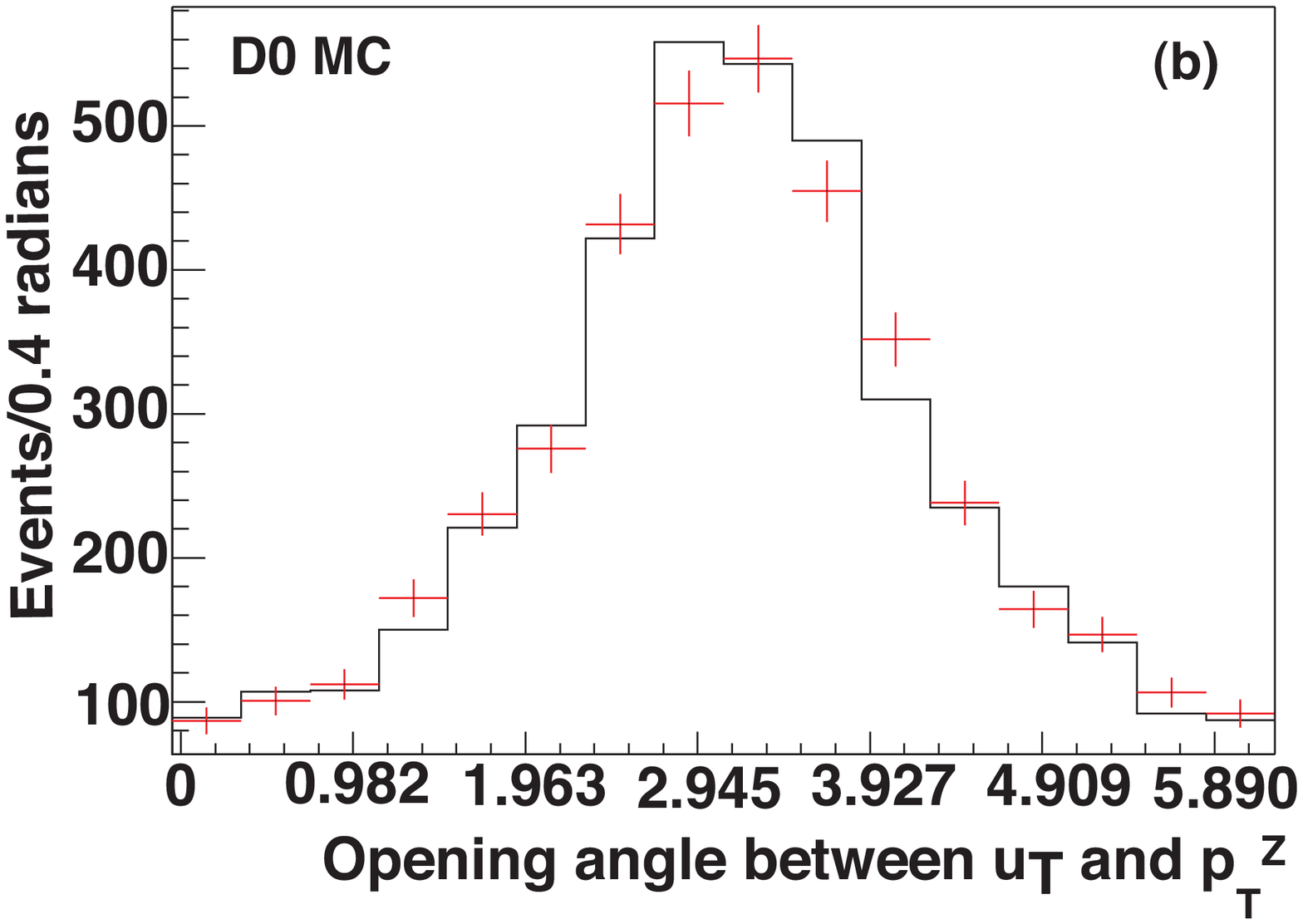}
  \caption{Opening angle between $\vec{u}_T$ and true $\vec{p}^{~Z}_T$ (solid line) and opening angle between $\vec{u}_T$ and the 
           estimated direction of true $\vec{p}^{~Z}_T$ (points) when using (a) the two smeared electrons directly and (b) the 
           unfolded map for $Z$ boson events with a true $|\vec{p}^{~Z}_T|$ of 4.0 to 4.25 GeV.}
  \label{fig:Rphi_unfolded}
\end{figure*}

Figure~\ref{fig:prob_dist} shows an example distribution of the probabilities that 
a $Z$ boson with a reconstructed $|\vec{p}^{~Z}_T|$ of 7 GeV and a recoil 
$|\vec{u}_T|$ of 3.5 GeV corresponds to different true $|\vec{p}^{~Z}_T|$ values. 
These probabilities are used to weight the given recoil 
as we store it in the recoil histograms corresponding to the true $|\vec{p}^{~Z}_T|$. 
Figures~\ref{fig:RpT_unfolded}--\ref{fig:Rphi_unfolded} show 
various recoil observables plotted versus the true $|\vec{p}^{~Z}_T|$, 
obtained from the truth information of these MC samples, compared with the same 
observables plotted versus the reconstructed $|\vec{p}^{~Z}_T|$, 
before and after the unfolding is applied. The unfolding 
corrects for average effects of $|\vec{p}^{~Z}_T|$ smearing on both the means 
and the RMS values of these recoil observables. 

\section{Uncertainties Particular to the Recoil Library Method}
\label{sec-rec-systs}
To perform high statistics tests of the efficacy of
the recoil library, we study the mass and width values
obtained by comparing $|\vec{p}_T^{~e}|$, $|\vmet|$ and $M_T$ distributions obtained from
fast MC $W$ boson samples created using the parameterized
recoil model with templates generated from $W$ boson samples 
created using the recoil library method. The recoil libraries are generated from
\zee ~events created with the parameterized recoil method. By varying
parameters in the simulation used to generate the $W$ boson samples
while leaving the templates unchanged, we measure the biases and 
statistical and systematic uncertainties on the recoil library method for 
$p\bar{p}$ collisions at $\sqrt{s}=1.96$ TeV. The corresponding uncertainties 
for $W$ boson mass and width measurements at the LHC remain to be evaluated,
but are not expected to be large.

\subsection{Statistical power of the $Z$ recoil sample}
\label{sec-rec-systs-stats}
There are significant statistical uncertainties since we obtain the 
recoil system for modeling the $\wen$ events from the limited sample of $Z$ boson events. 
In 1 fb$^{-1}$ of data, after the selection cuts, we expect approximately 
18,000 $Z\rightarrow ee$ events with both electrons in the central calorimeter, 
whereas in the same data we expect approximately 500,000 $\wen$ events with 
the electron in the central calorimeter. For the recoil library method, 
we choose recoil vectors from the same set of ~18,000 $Z\rightarrow ee$ events to make $W$ boson templates. 
Our method is thus limited by the size of the $Z$ recoil sample and any 
statistical fluctuations it contains. If we are to rely on this method 
as an input to a precision measurement, we need to determine the extent 
to which the statistical limitations of the $Z \rightarrow ee$ sample 
propagate to an uncertainty on the measured $W$ boson mass and width.

We assess the statistical uncertainties of the recoil method 
using an ensemble of 100 fast MC simulations resembling the statistical situation 
we expect in real data. We generate $W$ and $Z$ boson samples corresponding to 1 fb$^{-1}$ 
of data using the parameterized recoil method. For each set of $W$ and $Z$ boson samples, 
we use the $Z$ boson events to create a recoil library and then use the library to create templates 
for the recoil in the simulated $W$ boson events. These templates are then used to extract the 
$W$ boson mass and width. The statistical power is measured using the spread of extracted 
masses and widths from these ensembles. 

\begin{figure*}[htbp]
  \centering
    \includegraphics[scale=0.41]{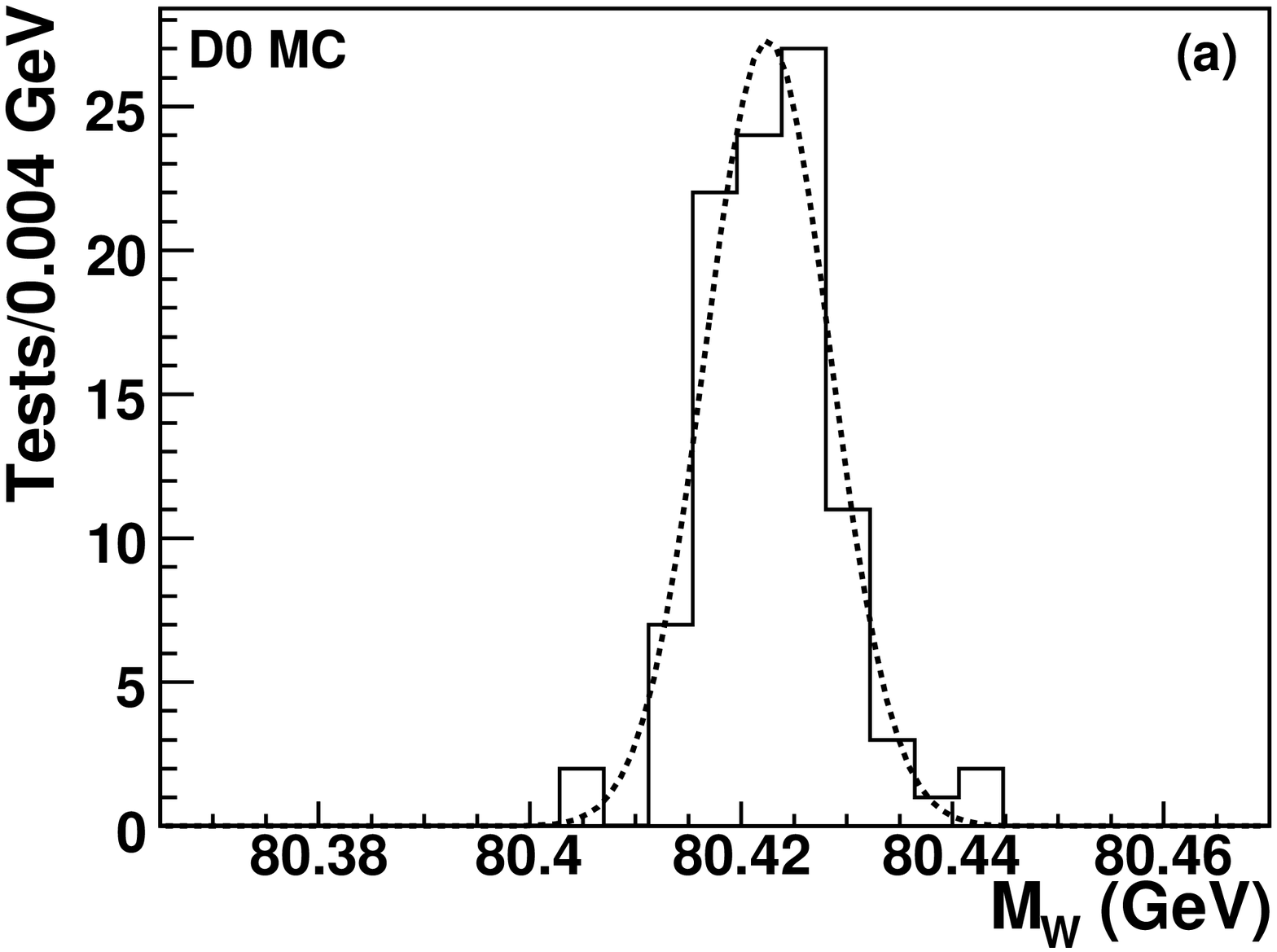}\hfil
    \includegraphics[scale=0.41]{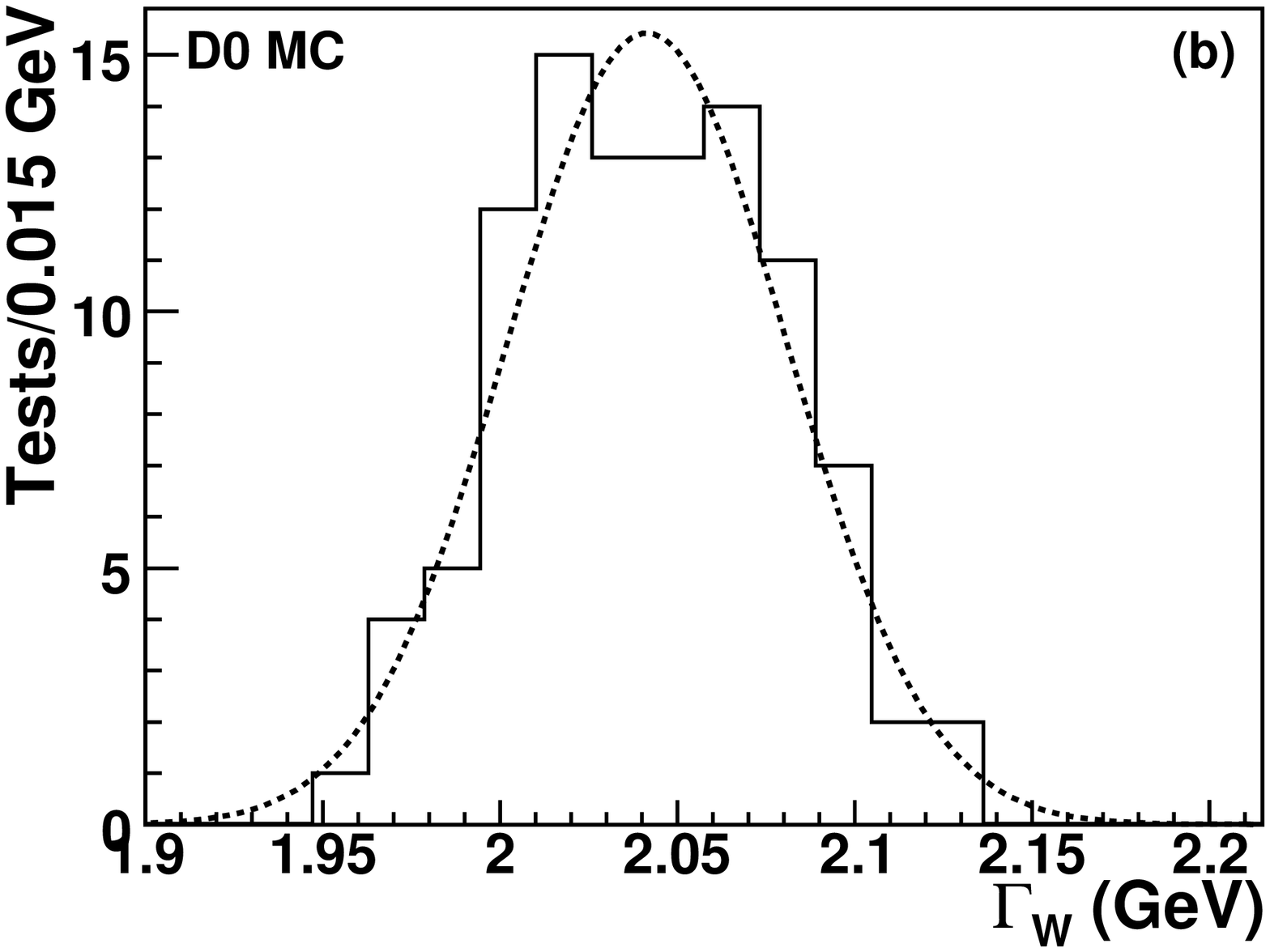}
  \caption{(a) $W$ boson mass and (b) width measured in 100 ensemble tests for each template generated from a recoil file.
    The dash line is a fit using a Gaussian function.
    All ensembles were generated with an input $W$ boson mass of 80.419 GeV and width of 2.039 GeV. 
    The fitted gaussian function for the mass has a mean value of $80.420 \pm 0.001$ GeV and RMS 
    of $0.005 \pm 0.001$ GeV. The values for the width are $2.040 \pm 0.001$ GeV (mean) and $0.040 \pm 0.003$ (RMS) GeV.}
  \label{fig:mass_stat_uncert}
\end{figure*}

Figure~\ref{fig:mass_stat_uncert} shows the measured $W$ boson masses and 
widths from 100 ensembles using the $M_T$ distribution. 
The mean fit value is in good agreement with the input value, showing that the recoil library can 
accurately model the parameterized recoil method. We test that the recoil library can model 
the full MC events in Section 6.

The statistical uncertainty on the mass measurement due to the recoil library method is found to be 
5 MeV from the $M_T$ spectrum, 8 MeV for the $|\vec{p}_T^{~e}|$ spectrum, and 17 MeV for 
the $|\vmet|$ spectrum. These agree with the statistical uncertainties on the 
parameterized recoil method, which are found to be 6 MeV for the $M_T$ fit, 7 MeV for 
the $|\vec{p}_T^{~e}|$ fit, and 19 MeV for the $|\vmet|$ fit. The statistical 
uncertainty on the width measurement due to the recoil library method is 40 MeV 
using the $M_T$ spectrum and agrees with the statistical uncertainty of 42 MeV 
using the parameterized recoil method. 

Both the parameterized recoil and the recoil library methods use the $Z$ boson sample to model the recoil.
One might naively expect that the additional information contained in the functional form used 
in the parameterized method would give it increased statistical power for the same-sized sample.
However, we do not observe a loss of statistical power since the uncertainties from the 
two methods are very similar to each other. We have 
explored the reason for this by using a simplified detector model of $W$ and $Z$ boson events 
with and without recoil energy resolution effects added, and comparing the $p_T$-imbalance 
(the difference between $|\vec{p}_T^{~Z}|$ and the projection of the recoil $\vec{u}_T$ along 
the boson direction) distribution for the parameterized and library methods. Due to the similar 
transverse momentum distributions of the $W$ and $Z$ bosons, we find that the means of the $p_T$-imbalance 
distribution agree with each other within statistical uncertainty. We also find that without recoil 
energy resolution effects, there is a clear but small, ${\cal{O}}(100)$ MeV, increase 
in the RMS of the $p_T$-imbalance distributions for the recoil library method, but with 
the detector resolution effects added, the RMS of the $p_T$-imbalance distribution increases 
to over 2 GeV and masks any difference stemming from the difference between the parameterized 
recoil method and the recoil library method.

\subsection{Systematic uncertainties}
\label{sec-rec-systs-effects}
We mentioned in Section~\ref{sec-method} that several 
effects could potentially bias the recoil library method.
These include unmerged FSR photons, acceptance differences 
between $Z$ and $W$ boson events, residual efficiency-related correlations between the electron
and the recoil system, underlying energy corrections 
beneath the electron window, and the unfolding process. 
The closure tests using fast MC described in Section~\ref{sec-rec-systs-stats} 
show the overall bias from this method to be smaller than the statistical 
power of the tests. Nonetheless, we want to make sure that this small 
final bias is not due to the cancellation of larger individual biases
and therefore examine each effect independently.

\subsubsection{Unmerged FSR photons}
\label{subsubsec-fsrbias}
We measure the residual bias due to FSR photons by fitting 
two sets of fast MC simulations, one made from an unfolded, high statistics recoil file 
with all FSR photons included and one made from an equivalent recoil file with no FSR photons. 
We find that the mass shift between these two samples is $-1$ MeV for the $M_T$ fit, 
$-2$ MeV for the $|\vec{p}_{T}^{~e}|$ fit, and 2 MeV for the $|\vmet|$ fit, and that 
the width shift is less than 1 MeV.

\subsubsection{Differences in geometric acceptance}
\label{subsubsec-accbias}
For $W$ candidates, we only require the electron to be in the central calorimeter, while 
for $Z$ candidates used to create the library, we require both electrons to be in the 
central calorimeter. To test the bias due to this effect, we generate two recoil 
files. For one recoil file we restrict both electrons to the central 
region of the detector, as we would in data. For the other recoil file, 
we restrict only one electron to the central calorimeter
and allow the other electron to be anywhere, as with the 
neutrino from the $W$ boson decay. We make templates from the two recoil 
files and find that the differences in both measured mass and measured width are 
smaller than the 2 MeV statistical uncertainty of this study. 

\subsubsection{Efficiency related biases}
\label{subsubsec-effbias}
When we generate unfolded recoil files, we weight 
the events by the reciprocals of the $u_{\parallel}^e$ and $SE_T$ efficiencies, 
as described in Section~\ref{subsec-efficiences}. 
To check if this approach introduces any biases, we perform three tests. 
For one, we only apply the $u_{\parallel}^e$ efficiency. 
In the second test, we only apply the $SE_T$ efficiency, and in the 
final test we apply both efficiencies. The maximum bias in the fitted mass 
or width over all three tests is used as the systematic uncertainty. 
The final uncertainty attributed to the efficiency corrections 
on the $W$ boson mass is 7 MeV for the $M_T$ fit, 7 MeV for the $|\vec{p}_T^{~e}|$ fit, and 8 MeV 
for the $|\vmet|$ fit. The uncertainty of the $W$ boson width is found to be 7 MeV.

\subsubsection{Uncertainty in $\Delta u^e_{\parallel}$}
In Section~\ref{sec-uparcorr} we observed that by removing 
the electrons from the $Z \rightarrow ee$ recoil file, we also 
remove the recoil energy that underlies the electron cones. 
We correct for this effect by adding back the average energy, 
$\Delta u^e_{\parallel}$, expected beneath the electrons. 
We then subtract the electron energy that leaks outside of the electron 
cone that is incorrectly attributed to the recoil energy.

We assess the systematic uncertainty due to these corrections as follows.
$Z$ boson recoil files are made for three cases: (1) no energy corrections,
(2) a constant energy correction for underlying hadronic energy beneath 
the electron cone and constant correction for the electron energy leakage,
(3) the parameterized energy correction for underlying hadronic 
energy beneath the electron cone and constant correction for the electron energy leakage.

We then generate three sets of templates from each of these recoil files and 
measure the shift in fitted $W$ boson mass and width between these three 
template sets. The $W$ boson mass shifts by 2 MeV for the $M_T$ fit, 4 MeV 
for the $|\vec{p}_T^{~e}|$ fit, and 1 MeV for the $|\vmet|$ fit, with a 7 
MeV shift for the width. We assign the magnitude of these maximum shifts 
as the uncertainty on this method due to the $\Delta u_{\parallel}^e$ correction.

\subsubsection{Uncertainties due to implementation of unfolding}
The specific choices made in implementing the unfolding could introduce 
biases to the final measurement. Our results may depend on our 
choice of initial distributions in $(|\vec{p}^{~Z}_T|)^t$, $\psi$, 
and $|\vec{u}_T|$. They could also depend on the number of iterations of the 
unfolding procedure we apply to the recoil library. 

We find that starting with flat initial distributions in 
$(|\vec{p}^{~Z}_T|)^t$, $\psi$, and $|\vec{u}_T|$, 
10 iterations are sufficient to attain convergence.
We generate the unfolded recoil files using 8, 10, and 12 iterations of the unfolding method and find that the 
changes in measured mass and width extracted from $M_T$, $|\vec{p}_T^{~e}|$, and $|\vmet|$ fits 
are within 3 MeV statistical uncertainty of the fast MC study. In addition to unfolding the recoil 
file using a flat initial distribution for the recoil spectrum, we also try several smoothly varying 
sinusoidal initial distributions, and find that the variation in the final 
unfolded recoil file is negligible.

\subsection{Total systematic uncertainties due to the recoil system simulation}
\label{subsubsec-toterr}
Table~\ref{tab:systmasswidth} shows the estimated systematic uncertainties 
due to the recoil system simulation for 1 fb$^{-1}$ of fast MC data. The overall 
systematic uncertainties, obtained by adding the individual 
uncertainties in quadrature, are found to be 9 MeV using the 
$M_T$ fit, 12 MeV using the $|\vec{p}_T^{~e}|$ fit, and 19 MeV 
using the $|\vmet|$ fit for the $W$ boson mass, and 41 MeV 
using the $M_T$ fit for the $W$ boson width.

\section{Full MC closure of $W$ boson mass and width}
\label{sec-fullMC}
We test both the recoil library method and the parameterized recoil method using a 
detailed MC sample produced using a {\sc geant}-based full detector model 
for $W$ and $Z$ boson production. The full MC $Z$ boson sample is 
equivalent to 6.0 fb$^{-1}$ and the $W$ boson sample is equivalent to 2.5 fb$^{-1}$. 
In this case, the full MC $Z$ boson samples are used to 
create the recoil library. Templates are then created from $W$ boson samples
made using the recoil library, and these are used to extract the $W$ boson mass
and width. The extracted values for the $W$ boson mass and width are compared to 
the input values (closure test).

\begin{figure*}[htbp]
\centering
\includegraphics [scale=0.34] {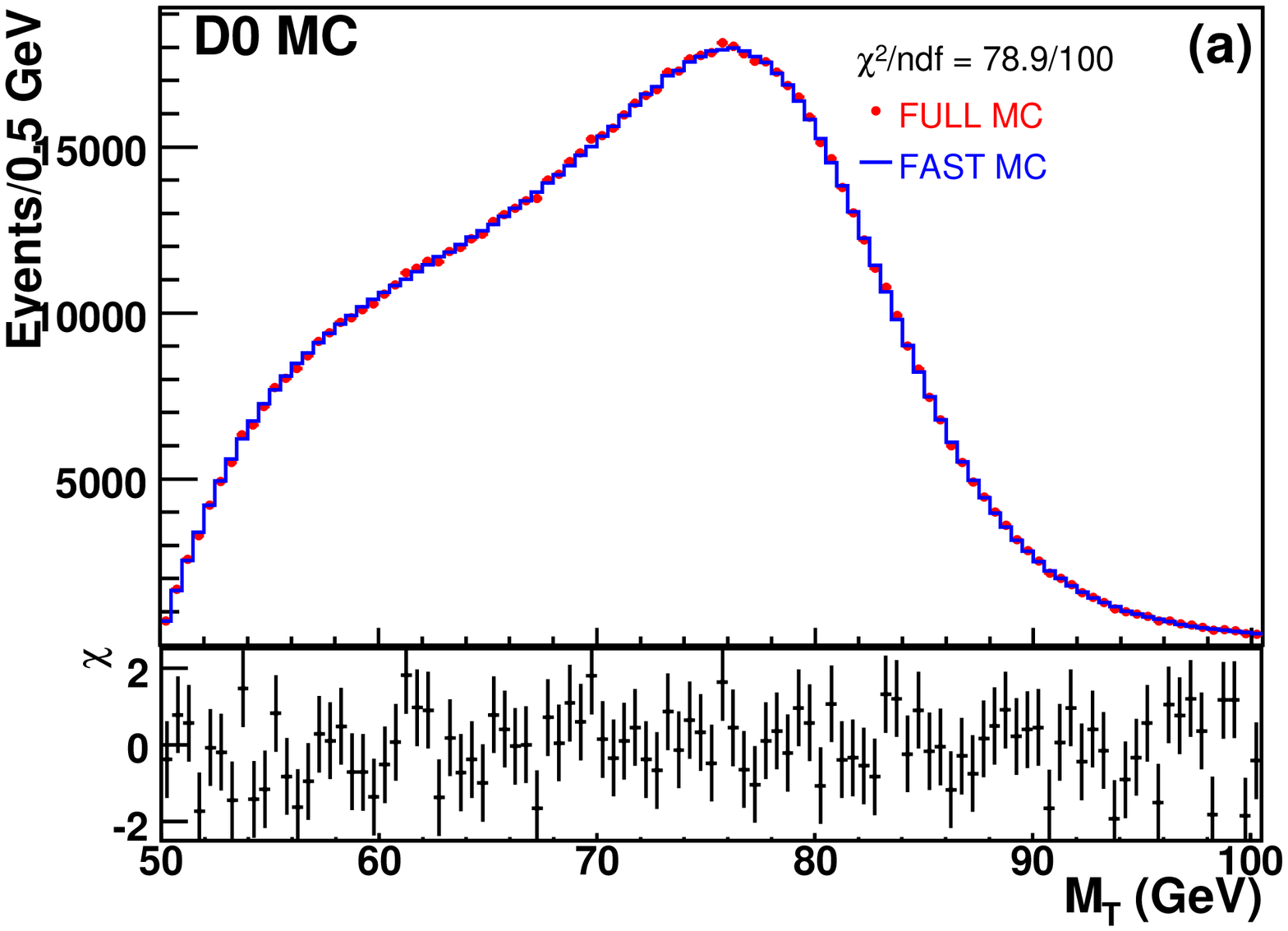} 
\includegraphics [scale=0.34] {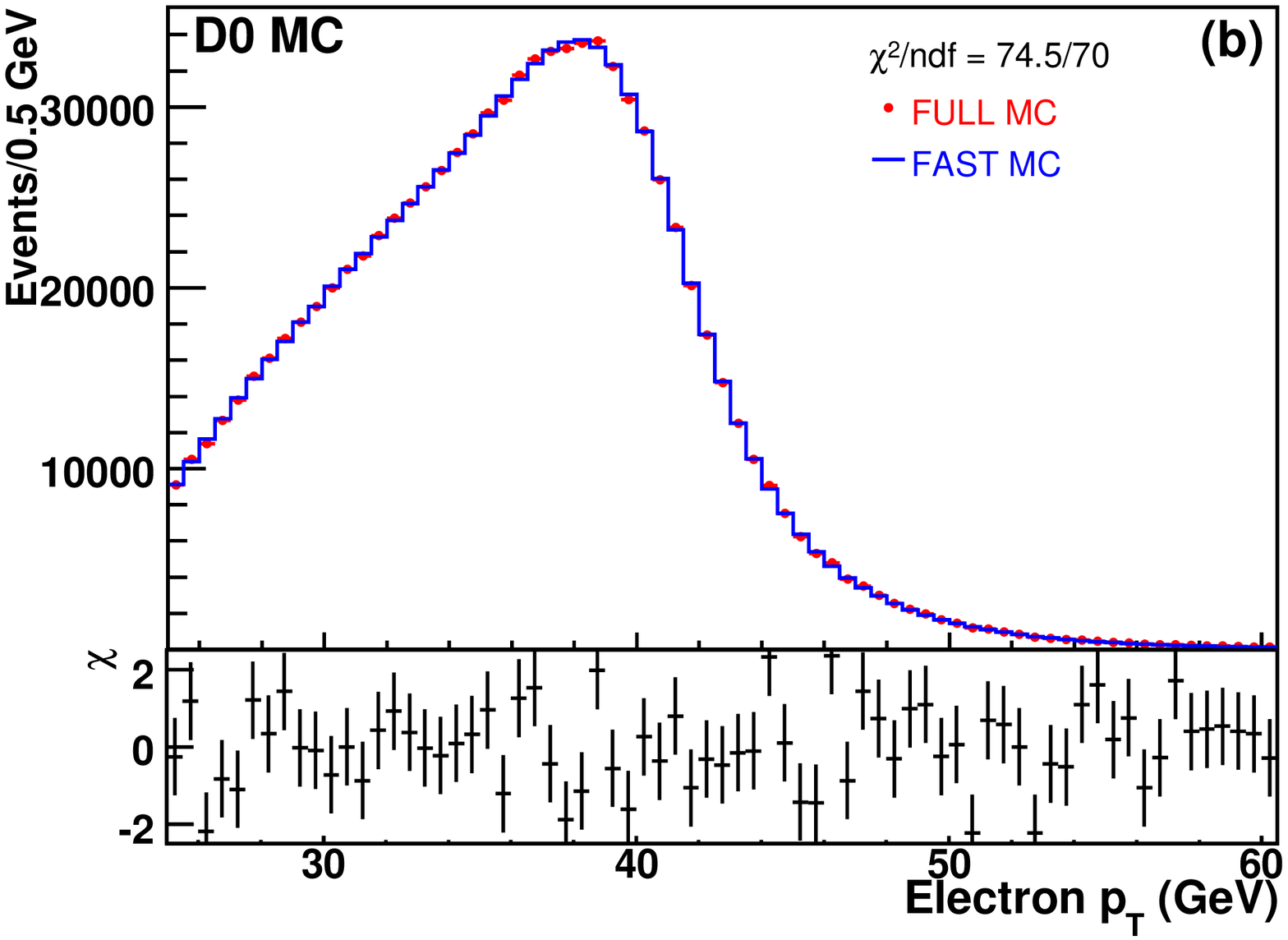} 
\includegraphics [scale=0.34] {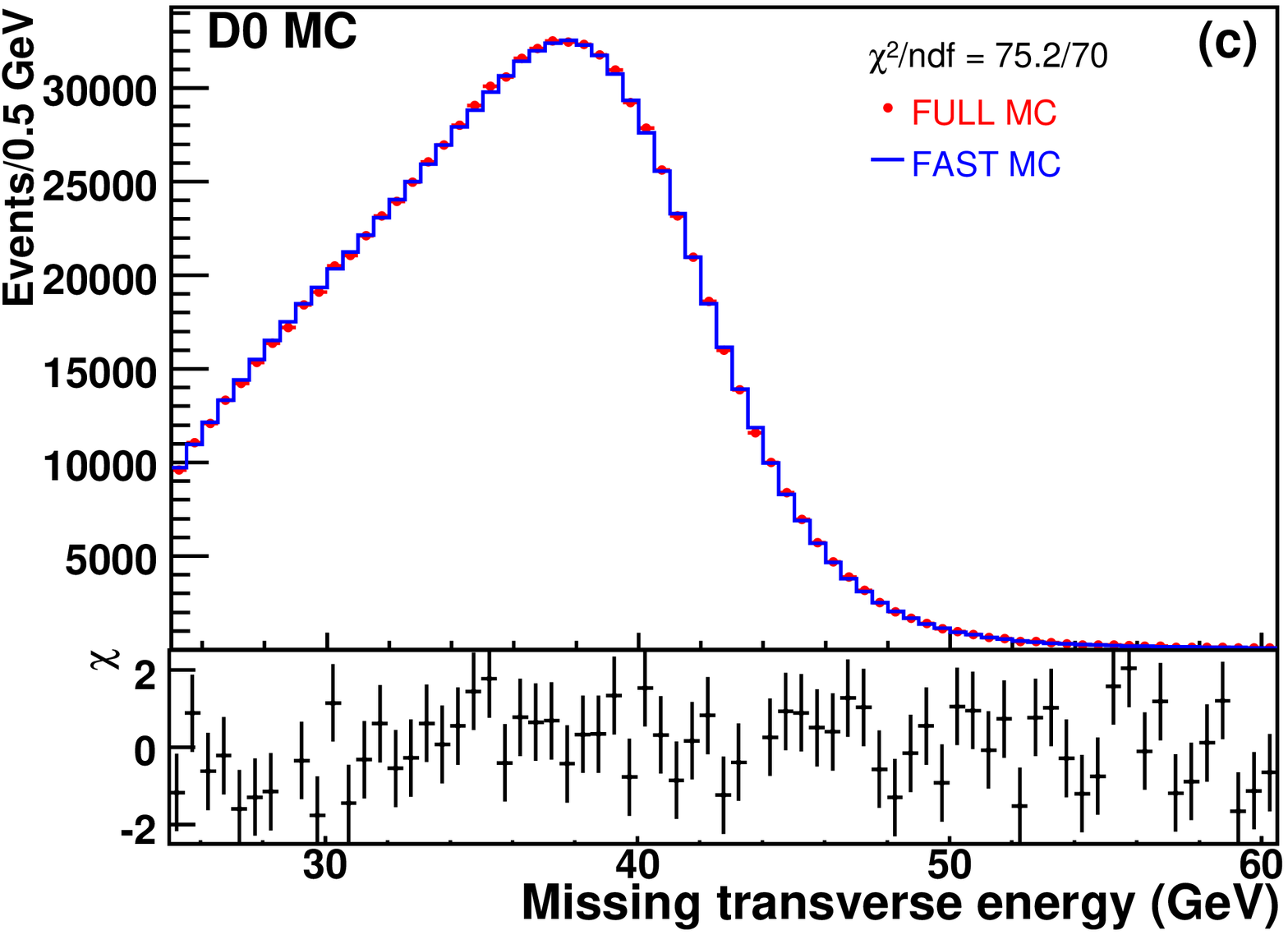} 
\includegraphics [scale=0.34] {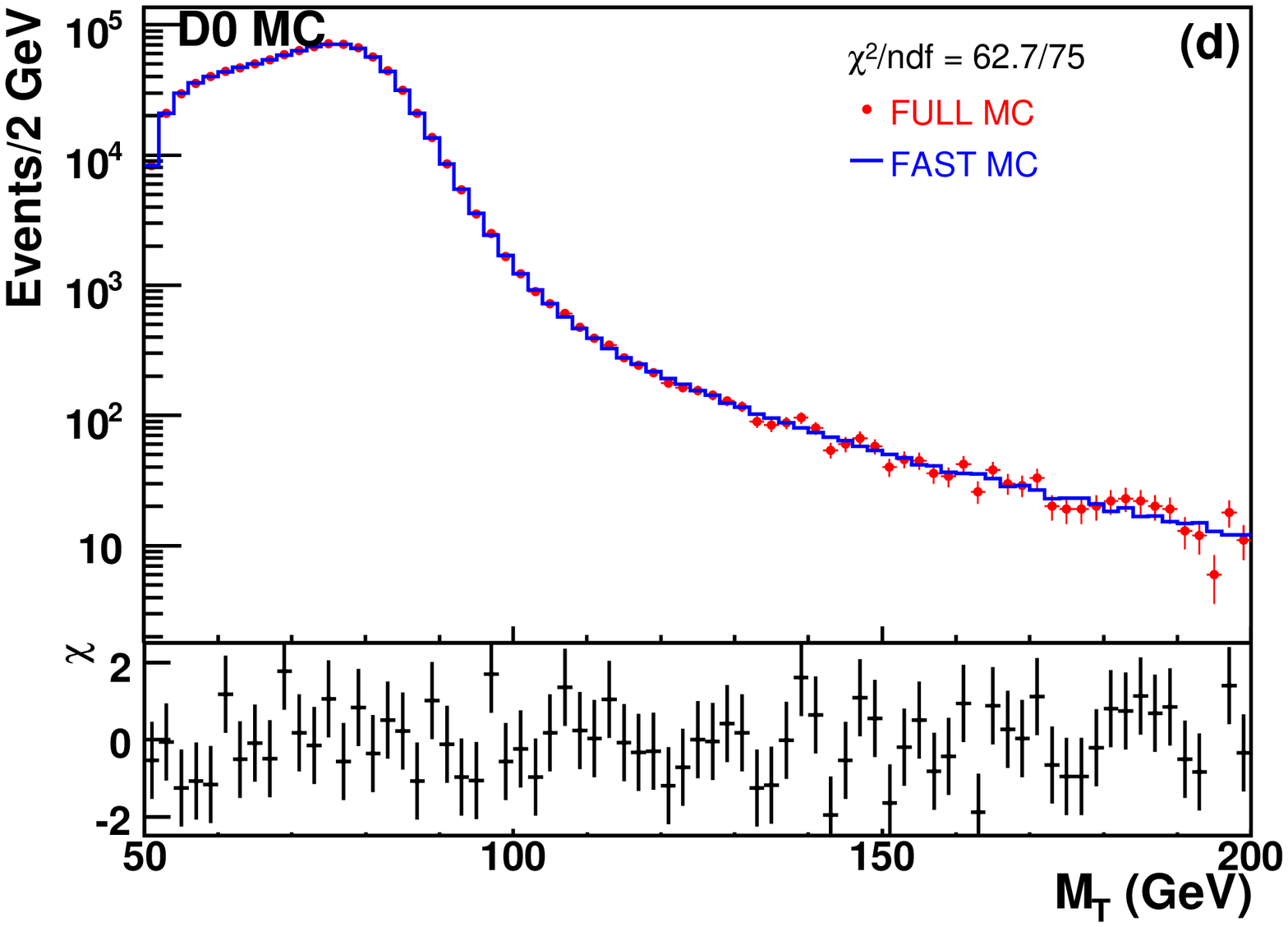} 
\caption{Comparison plots between full MC (points) and fast MC produced using the recoil library (lines) 
for the $W$ boson (a) $M_T$, (b) $|\vec{p}_T^{~e}|$, (c) $|\vmet|$, and (d) $M_T$ (log scale) distributions. 
Also shown are the $\chi$ values defined as the difference between full MC and fast MC yields divided 
by the statistical uncertainty on the full MC yield. Different ranges and bin sizes are used for (a) and (d).}
\label{fig:full_MC_maincomp}
\end{figure*}

Before fitting for the mass and width of the full MC sample, 
we test the accuracy of the model by comparing various full 
MC distributions to the fast MC model for an input value of the $W$ boson mass of 80.450 GeV. 
Good agreement between full MC and fast MC using the recoil library method is observed.
Figure~\ref{fig:full_MC_maincomp} shows comparisons between 
$W \rightarrow e \nu$ full MC and fast MC using the recoil library method for the 
$M_T$, $|\vec{p}_{T}^{~e}|$, and $|\vmet|$ distributions.
The $\chi^2$ between full and fast MC simulations are also given and are reasonable. 
The systematic uncertainties on the electron model, dominated 
by the uncertainty on the electron energy scale, are found to be 15 MeV for the $M_T$ and $|\vmet|$ fits, and
12 MeV for the $|\vec{p}_T^{~e}|$ fit for the $W$ boson mass, and 15 MeV for the $W$ boson width.
Systematic uncertainties on the hadronic model are 
taken from Section V. Since here we use the equivalent of 6.0 fb$^{-1}$ of full MC 
$\zee$ recoils, we estimate the overall uncertainty due to the recoil system 
simulation by scaling the uncertainty due to recoil statistics found in Section V by a factor 
of $1/\sqrt{6}$, leaving other estimated systematic uncertainties the same.
The systematic uncertainty due to the recoil statistics is 2 MeV using the 
$M_T$ fit, 3 MeV using the $|\vec{p}_T^{~e}|$ fit, and 7 MeV using the $|\vmet|$ fit for the 
$W$ boson mass and 16 MeV using the $M_T$ fit for the $W$ boson width, which agrees with 
the corresponding systematic uncertainty in the parameterized recoil model. Taking the 
systematic uncertainties estimated in Section V, added in quadrature with these statistical 
uncertainties, we find the total uncertainty to be 22 MeV for the $M_T$ fit, 24 MeV for the 
$|\vec{p}_T^{~e}|$ fit and 26 MeV for the $|\vmet|$ fit for the $W$ boson mass, and 36 MeV for the $W$ boson width.

The results of the full MC measurements agree 
with the full MC input $W$ boson mass and width values within the uncertainties,
as shown in Table~\ref{tab:closure_mass}. 

\section{Conclusion}
\label{sec-conclusion}
We have outlined a method to model the hadronic recoil system in $W \rightarrow \ell \nu$ 
events using recoils extracted directly from a $Z \rightarrow \ell \ell$ data library. We applied this 
methodology to a realistic full MC simulation of the D0 detector. 
The $W$ boson mass and width fits to these MC events are in good agreement with the input parameters, 
within statistical uncertainties. They also agree with the values extracted using a more traditional parameterized 
recoil model. Comparisons of simulated distributions using the recoil library method with MC give 
good $\chi^{2}$ agreement over a full range of data observables. 

This method is limited by the statistical power of the $Z$ boson recoil sample, as is the parameterized recoil 
model. In addition to systematic effects from the limited statistical power of the $Z$ boson sample, 
there are several systematic effects due to the implementation of the selection efficiencies, geometric acceptance, 
the unfolding method, and FSR. The uncertainty due to these effects is found to be ${\cal{O}}(10)$ MeV.  

The method presented in this paper has many advantages. It accurately describes the highly 
complicated hadronic response and resolution for $W$ boson recoils in a given calorimeter. It 
includes complex correlations between the hard and soft components of the recoil
and scales the recoil appropriately with luminosity. It requires fewer assumptions, no 
first-principles description of the recoil system, and no adjustable parameters. 
At hadron collider experiments at the Run II Tevatron and the LHC, this approach 
to modeling the recoil system is complementary to the traditional parametric approach.

\section*{Acknowledgement}
%
We thank the staffs at Fermilab and collaborating institutions, 
and acknowledge support from the 
DOE and NSF (USA);
CEA and CNRS/IN2P3 (France);
FASI, Rosatom and RFBR (Russia);
CNPq, FAPERJ, FAPESP and FUNDUNESP (Brazil);
DAE and DST (India);
Colciencias (Colombia);
CONACyT (Mexico);
KRF and KOSEF (Korea);
CONICET and UBACyT (Argentina);
FOM (The Netherlands);
STFC and the Royal Society (United Kingdom);
MSMT and GACR (Czech Republic);
CRC Program, CFI, NSERC and WestGrid Project (Canada);
BMBF and DFG (Germany);
SFI (Ireland);
The Swedish Research Council (Sweden);
CAS and CNSF (China);
and the
Alexander von Humboldt Foundation (Germany).

\bibliographystyle{elsarticle-num}
\bibliography{paper}

\begin{thebibliography}{00}
\bibitem{bib:geo}
D0 uses a cylindrical coordinate system with the $z$ axis directed
along the beam axis in the proton direction. Angles $\theta$ and $\phi$ are the polar and
azimuthal angles, respectively. Pseudorapidity is defined as
$\eta=-\ln [\tan(\theta/2)]$ where $\theta$ is measured with respect
to the interaction vertex. In the massless limit, $\eta$ is equivalent to
the rapidity $y=(1/2) \ln[(E+p_z)/(E-p_z)]$.
$\eta_{D}$ is the pseudorapidity measured with respect to the center of the detector.
\end{thebibliography}


\clearpage

\clearpage
\begin{table*}[htbp]
\begin{center}
\caption{Total systematic uncertainties on the $W$ boson mass and width 
from the recoil library method, for 1 fb$^{-1}$ of $Z$ boson data.} 
\begin{tabular}{lcccc}
\hline
Source&$\Delta M_W$($M_T$) & $\Delta M_W$($|\vec{p}_T^{~e}|$) & $\Delta M_W$($|\vmet|$) & $\Delta \Gamma_W$($M_T$) \\ 
    & (MeV) & (MeV) & (MeV) & (MeV) \\
\hline
Recoil statistics &5&8&17&40\\
FSR photons &1&2&2&1\\
Efficiency related bias&7&7&8&7\\
$\Delta u^e_{\parallel}$&2&4&1&7\\
Unfolding&3&3&3&3\\
\hline
Systematic total &9&12&19&41\\
\hline
\end{tabular}
\label{tab:systmasswidth} 
\end{center}
\end{table*}

\begin{table*}[htbp]
\begin{center}
\caption{Final result of the full MC closure fits for the $W$ boson mass and width using the 
recoil library method. The full MC samples used here are equivalent to 2.5 fb$^{-1}$ of $W$ 
boson data and 6.0 fb$^{-1}$ of $Z$ boson data. For the fitted $W$ boson mass and width, the 
first uncertainty is statistical, the second is the systematic on the electron simulation, 
the third is the systematic on the recoil system simulation due to $Z$ boson statistics, and the 
fourth is other systematics on the recoil system simulation. $\Delta M_W$ represents the 
difference between the measured $W$ boson  mass and the input value of 80.450 GeV, and $\Delta \Gamma_W$ 
represents the difference between the measured $W$ boson width and the input value of 2.071 GeV.}
\begin{tabular}{cc}
\hline
Observable & $\Delta M_W$ (MeV)\\
\hline
$M_T$& 6 $\pm$ 15 $\pm$ 15 $\pm$ 2 $\pm$ 7 \\
$|\vec{p}_T^{~e}|$& 5 $\pm$ 19 $\pm$ 12 $\pm$ 3 $\pm$ 8 \\
$|\vmet|$& 0 $\pm$ 19 $\pm$ 15 $\pm$ 7 $\pm$ 8 \\
\hline
 & $\Delta \Gamma_W$ (MeV) \\
\hline
$M_T$& $-5$ $\pm$ 27 $\pm$ 15 $\pm$ 16 $\pm$ 10 \\
\hline
\end{tabular}
\label{tab:closure_mass} 
\end{center}
\end{table*}

\end{document}